\documentclass{autart}
\usepackage[english]{babel}
\usepackage{amsmath}
\usepackage{amsfonts}
\usepackage{amssymb}
\usepackage{graphicx}

\newcommand{\Tr} {\mbox{\rm tr}}

\newcommand{\diag} {\mbox{\rm diag}}

\begin{document}

\begin{frontmatter}

\title{Robust EM kernel-based methods\\ for linear system identification$^\star$ } % Title, preferably not more than 10 words.

\thanks[footnoteinfo]{The research leading to these results has received funding
from the Swedish Research Council under contract 621-2009-4017,
the European Union Seventh Framework Programme
[FP7/2007-2013] under grant agreement no. 257462 HYCON2 Network of excellence, by the MIUR
FIRB project RBFR12M3AC - Learning meets time:
a new computational approach to learning in dynamic
systems, and by the Washington Research Foundation Data Science Professorship.}

\author[First]{Giulio Bottegal}
\author[Second]{Aleksandr Y. Aravkin}
\author[First]{H\r akan Hjalmarsson}
\author[Third]{Gianluigi Pillonetto}

\address[First]{Automatic Control Lab and ACCESS Linnaeus Centre, School of Electrical Engineering,
KTH Royal Institute of Technology, Stockholm, Sweden (e-mails: \{bottegal; hjalmars\}@kth.se)}
\address[Second]{Department of Applied Mathematics, University of Washington, Seattle, WA, USA   (e-mail: saravkin@uw.edu)}
\address[Third]{Department of Information Engineering, University of Padova, Padova, Italy  (e-mail: giapi@dei.unipd.it)}

                             % keyword list or with the
                                          % help of the Automatica
                                          % keyword wizard

\begin{abstract}                          % Abstract of not more than 250 words.
Recent developments in system identification have brought attention to regularized kernel-based methods.
This type of approach has been proven to compare favorably with classic parametric methods.
However, current formulations are not robust with respect to outliers.
In this paper, we introduce a novel method to robustify kernel-based system identification methods.
To this end, we model the output measurement noise using random variables with heavy-tailed probability density functions (pdfs),
focusing on the Laplacian and the Student's t distributions.
Exploiting the representation of these pdfs as scale mixtures of Gaussians,
we cast our system identification problem into a Gaussian process regression framework,
which requires estimating a number of hyperparameters of the data size order.
To overcome this difficulty, we design a new maximum a posteriori (MAP) estimator of the hyperparameters, and solve the related optimization problem with a novel iterative scheme based on the Expectation-Maximization (EM) method.
In presence of outliers, tests on simulated data and on a real system show a substantial performance improvement
compared to currently used kernel-based methods for linear system identification.

\end{abstract}

\end{frontmatter}

\section{Introduction} \label{sec:introduction}
Regularization techniques for linear regression have a very long history in statistics and data analysis \cite{Tikhonov}, \cite{Poggio}, \cite{Wahba1990}. Recently, in a series of papers, new regularization strategies have been proposed for
linear dynamic system identification \cite{SS2010}, \cite{SS2011}, \cite{ChenOL12}, \cite{pillonetto2014kernel}.
The basic idea is to define a nonparametric estimator of the impulse response of the system.
As compared to classic parametric methods such as the prediction error method (PEM)  \cite{Ljung}, \cite{Soderstrom},
the main motivation for this alternative approach is to avoid the model order selection step, which is usually required in parametric methods.
If no information about the structure of the system is given, in order to establish the number of parameters required
to describe the dynamics of the system, one has to rely on complexity criteria
such as AIC and BIC \cite{Akaike1974}, \cite{schwarz1978estimating} or cross validation \cite{Ljung}.
However, results of these criteria may not be satisfactory when only  short data sets are available \cite{PitfallsCV12}.

To circumvent model order selection issues, one can use regularized least-squares that avoid high variance in the estimates using a regularization matrix, related to the so called \emph{kernels} introduced in the machine learning literature.
In the context of system identification, several kernels have been proposed, e.g. the \emph{tuned/correlated} (TC), \emph{diagonal/correlated} (DC) kernels \cite{ChenOL12}, \cite{chen2014constructive},
and the family of \emph{stable spline kernels} \cite{SS2010,PillACC2010}.
Stable spline kernels have also been employed for estimating autocorrelation functions of stationary stochastic processes \cite{bottegal2013regularized}.

\begin{figure*}[!ht]
\begin{center}
\begin{tabular}{cc}
    {\includegraphics[width=8.5cm]{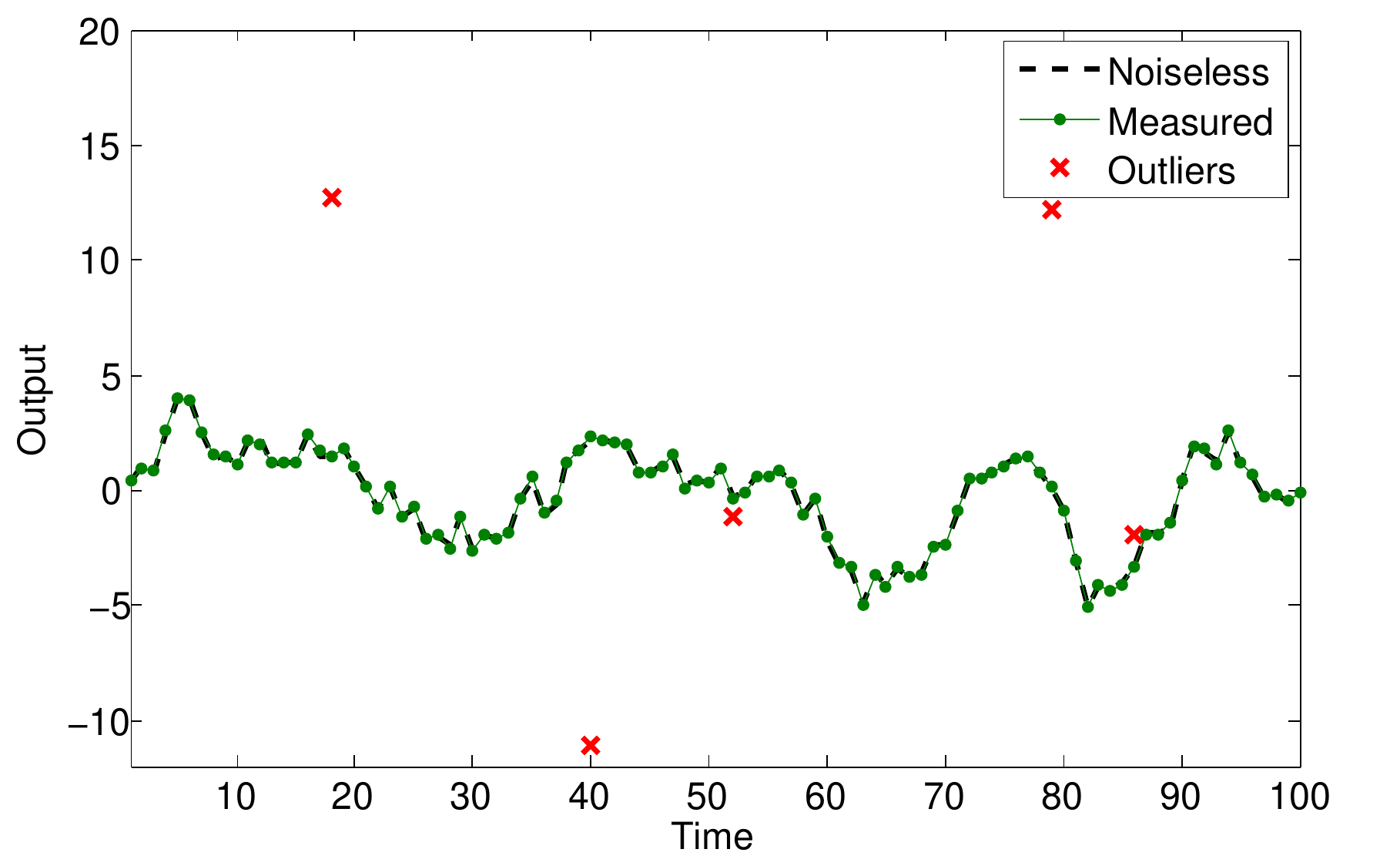}}
&
    {\includegraphics[width=8.5cm]{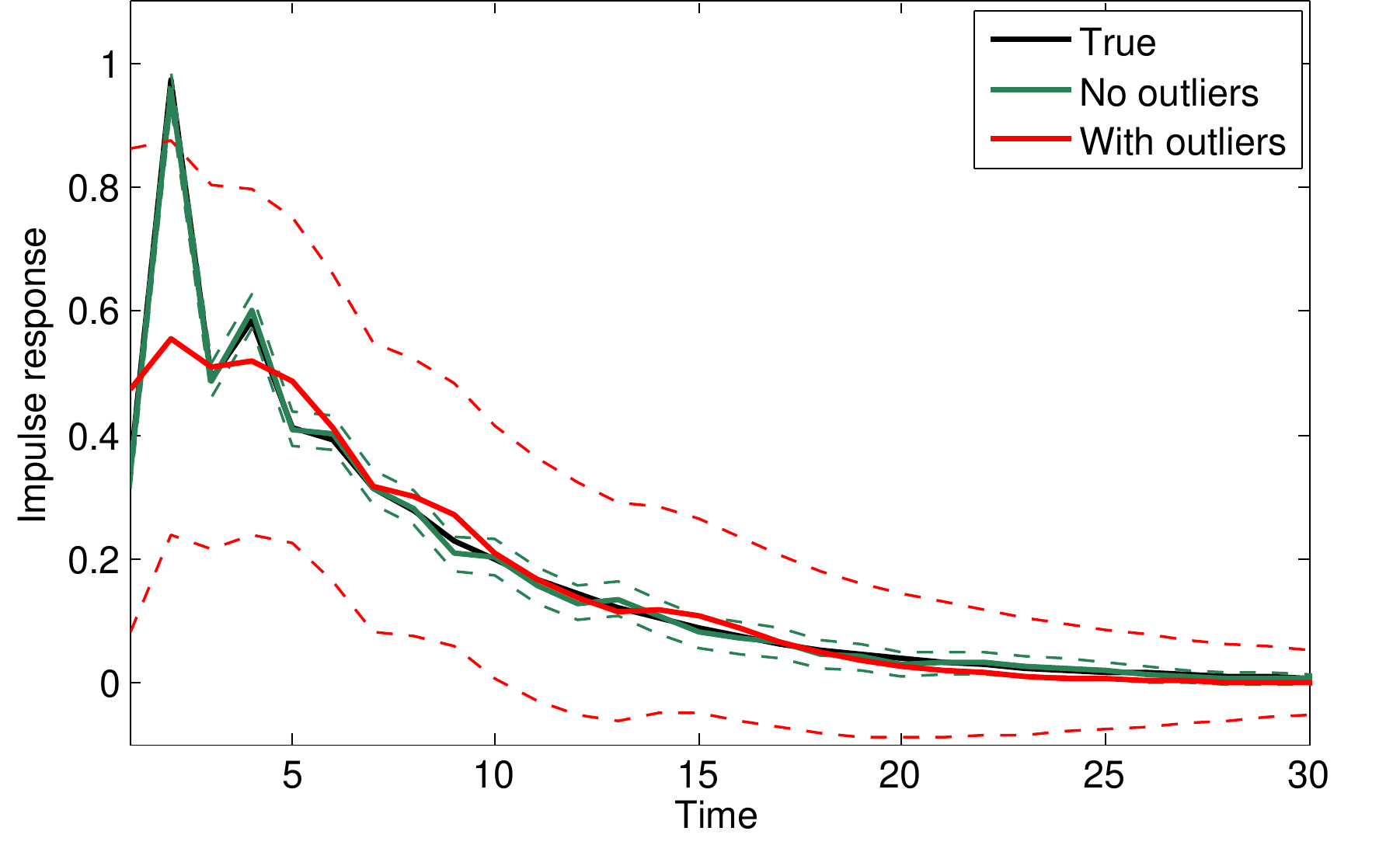}} \\
\end{tabular}
    \caption{\textbf{Introductory example.} Left panel: the noiseless output and the measured outputs in the no-outliers situation (measurements shown using green dots) and when outliers are present (shown using red asterisks). Right panel: the true impulse response and its estimate in the no-outliers situation and when outliers are present (the dashed lines represent the $99\%$ credibility bounds).} \label{fig:example}
\end{center}
\end{figure*}

In order to guarantee flexibility of regularized kernel-based methods,
the kernel structure always depends on a few parameters (in this context usually called \emph{hyperparameters}),
which are selected using available data; this process can be seen as the counterpart to model order selection in parametric approaches.
An effective technique for hyperparameter selection is based on the {\it Empirical Bayes} method \cite{Maritz:1989}.
Exploiting the Bayesian interpretation of regularization \cite{Wahba1990},
the impulse response is modeled as a Gaussian process whose covariance matrix corresponds to the kernel.
Hyperparameters are chosen by maximizing the marginal likelihood of the output data,
obtained by integrating out the dependence on the impulse response, see e.g. \cite{ChenML,TippingFast} for
algorithms for marginal likelihood optimization in system identification and Gaussian regression contexts.
Then, the unknown impulse response is retrieved by computing its minimum mean square error Bayesian estimate.
However, this approach, by relying on Gaussian noise assumptions, uses a quadratic loss to measure adherence to experimental data.
As a result, it can be non-robust when outliers corrupt the output data \cite{Aravkin2011tac}, as described in the following example.

\subsection{A motivating example} \label{sec:example}

Suppose we want to estimate the impulse response of a linear system fed by white noise using the kernel-based method proposed in \cite{SS2010}. We consider two different situations, depicted in Figure \ref{fig:example}. In the first one, 100 samples of the output signal are measured with a low-variance Gaussian additive noise (left panel);
note that the estimated impulse response is very close to the truth (right panel).
In the second situation we introduce 5 outliers in the measured output,
obtaining a much poorer estimate of the same impulse response.
This suggests that outliers may have a significant detrimental effect if kernel-based methods are not adequately robustified.

\subsection{Statement of contribution and organization of the paper}
%Motivated by the need of affording situations where the output measurements may be corrupted by outlier measurements,
We derive a novel outlier-robust system identification algorithm for linear dynamic systems.
The starting point of our approach is to establish a Bayesian setting
where the impulse response is modeled as a Gaussian random vector.
The covariance matrix of such a vector is given by the stable spline kernel,
which encodes information on the BIBO stability of the unknown system.

To handle outliers, we model noise using independent identically
distributed random variables with heavy-tailed probability densities.
Specifically, we make use of the Laplacian and the Student's t distributions.
%Approaches based on non-Gaussian noise models have been proposed in system identification, see e.g. \cite{dahlin2012hierarchical}, \cite{bottegal2014outlier}.
In order to obtain an effective system identification procedure, we
exploit the representation of these noise distributions as scale mixtures of Gaussians \cite{andrews1974}.
Each noise sample is seen as a Gaussian variable whose variance is unknown but has a prior distribution
that depends on the choice of the noise distribution.
The variance of each noise sample needs to be estimated from data.
To accomplish this task, we propose a novel maximum a posteriori (MAP) estimator
able to determine noise variances and kernel hyperparameters simultaneously.
Making use of the Expectation-Maximization (EM) method \cite{dempster1977EM,VarBayesEM},
we derive a new iterative identification procedure based on a very efficient
joint update of all the optimization variables. % is very efficient admit closed-form expressions for their updates, with the exception of one of the kernel hyperparameters.
The performance of the proposed algorithm is evaluated by numerical experiments.
When outliers corrupt the output data, results show that there is a clear advantage of the new method compared to
the kernel-based methods proposed in \cite{SS2010} and \cite{ChenOL12}. This evidence is supported by an experiment on a real system, where the collected data are corrupted by outliers.

It is worth stressing that robust estimation is a classic and well studied topic in applied statistics and data analysis.
Popular methods for robust regression hinge on the so-called M-estimators (such as the Huber estimator) \cite{huber2011robust}
or on outlier diagnostics techniques \cite{rousseeuw2005robust}. In the context of Gaussian regression,
recent contributions that exploit Student's t noise models can be found also in \cite{TippingSt,VanSt}.
In the system identification context, some outlier robust methods have been developed in recent years
\cite{pearson2002outliers,rojo2004support,Aravkin2011tac,sadigh2014robust}.
In particular, \cite{dahlin2012hierarchical,bottegal2014outlier} use non-Gaussian descriptions of noise,
while \cite{AravkinCDC2013} describes a computational framework based on interior point methods.
In comparison with all these papers, the novelty of this work is to combine
kernel-based approaches, noise mixture representations and EM techniques,
to derive a new efficient estimator of the impulse response and kernel/noise hyperparameters.
In particular, we will show that the MAP estimator of the hyperparameters can be implemented
solving a sequence of one-dimensional optimization problems,
defined by two key quantities which will be called
{\it{a posteriori total residual energy}}
and {\it{a posteriori differential impulse response energy}}.
All of these subproblems, except one involving a parameter connected with the dominant pole of the system,
can be solved in parallel and admit a closed-form solution. This makes the proposed method computationally attractive, especially when compared to possible alternative solutions such  as approximate Bayesian methods (e.g., Expectation Propagation \cite{minka2001expectation} and Variational Bayes \cite{beal2003variational}), or full Bayesian methods \cite{bottegal2014outlier}.
%In particular, let $n$ be the
%number of unknown impulse response coefficients while $N$ is the data set size.
%Then, at every step, the subproblems are $N+2$, defined
%just inverting a matrix of size $n \times n$ (typically, $n \ll N$ in system identification).

%A Common methods to handle outliers both in identification and filtering problems are based either on adding a $\ell_1$ penalty to a certain cost function related to the estimation method \cite{sadigh2014robust}, \cite{Aravkin2011tac}, or on pre- or post-processing of data \cite{pearson2002outliers}. In our recent work \cite{bottegal2014outlier}, we proposed a Bayesian method based on Markov Chian Monte Carlo techniques. Compared to that method, the algorithm proposed in this paper shows a strongly reduced computational load.

%SVM for robust parameter identification \cite{rojo2004support}

%ARX with Student's t \cite{dahlin2012hierarchical}

The organization of the paper is as follows. In Section~\ref{sec:problem}, we introduce the problem of linear dynamic system identification.
In Section \ref{sec:bayesian}, we give our Bayesian description of the problem.
In Section \ref{sec:EM}, we introduce the MAP-based approach to the problem and describe how to efficiently handle it.
Numerical simulations and a real experiment to evaluate the proposed approach are in Section \ref{sec:numerical}. Some conclusions end the paper
while the Appendix gathers the proof of the main results.

\section{Problem statement}
\label{sec:problem}
\begin{figure}[!ht]
\begin{center}
{\includegraphics[width=5cm]{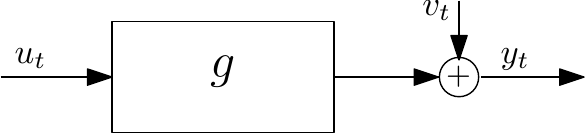}}
 \caption{Block scheme of the system identification scenario.} \label{fig:block_scheme}
\end{center}
\end{figure}

We consider a SISO linear time-invariant discrete-time dynamic system (see Figure \ref{fig:block_scheme})
\begin{equation} \label{eq:sys1}
y_t = \sum_{i=0}^{+\infty} g_i u_{t-i} + v_t \,,
\end{equation}
where $\{g_t\}_{t=0}^{+\infty}$ is a strictly causal transfer function (i.e., $g_0 = 0$) representing the dynamics of the system, driven by the input $u_t$. The measurements of the output $y_t$ are corrupted by the process $v_t$, which is zero-mean white noise with variance $\sigma^2$.
For the sake of simplicity, we will also hereby assume that the system is at rest until $t=0$.

We assume that $N$ samples of the input and output measurements are collected, and denote them by $\{u_t\}_{t=0}^{N-1}$, $\{y_t\}_{t=1}^{N}$.
Our system identification problem is to obtain an estimate of the impulse response $g_t$ for $n$ time instants,
namely $\{g_t\}_{t=1}^{n}$ ($g_0 = 0$).
\begin{rem}
The identification approach adopted in this paper allows setting $n = +\infty$ (at least theoretically), independently of the available data set size. For computational speed we will consider $n$ as \emph{large enough to capture the system dynamics}, meaning that $g_{n+1} \simeq 0$, i.e., the unmodeled tail of the impulse response is approximately zero (see also \cite{ChenOL12} for further details on the implications of this approximation). Recall that, by choosing $n$ sufficiently large, we can model $\{g_t\}_{t=1}^{+\infty}$ with arbitrary accuracy \cite{ljung1992}.
\end{rem}

Introducing the vector notation
$$
y := \begin{bmatrix} y_1 \\ \vdots \\ y_N \end{bmatrix} \,,\, g := \begin{bmatrix} g_1 \\ \vdots \\ g_n \end{bmatrix} ,\, v := \begin{bmatrix} v_1 \\ \vdots \\ v_N \end{bmatrix}
$$
%$$
%U = \begin{bmatrix} u_0 & 0  & & \ldots & 0 \\ u_1 & u_0 & 0 & \ldots & 0 \\ \vdots & \vdots &  &\ddots & \vdots \\ u_{N-2} & u_{N-3}  &  \ldots & u_{N-n+1}& 0 \\ u_{N-1} & u_{N-2}  &  \ldots &\ldots &  u_{N-n}  \end{bmatrix} \, \in \, \mathbb{R}^{N \times n}  \,,
%$$
and defining $U$ as the $N \times n$ Toeplitz matrix of the input (with null initial conditions), the input-output relation for the available samples can be written
\begin{equation} \label{eq:sys2}
y = Ug + v  \,,
\end{equation}
so that our estimation problem can be cast as a linear regression problem. We shall not make any specific requirement on the input sequence (i.e., we do not assume any  condition on persistent excitation in the input \cite{Ljung}), requiring only $u_t \neq 0$ for some $t$.
%However, well-posed input sequences (in the sense that all the singular values of $U^TU$ are bounded away from zero, see \cite{bertero1988ill}) are preferable, since they help in providing more accurate estimates of $g$.
\begin{rem}
The identification method we  propose in this paper can be derived also in the continuous-time setting, using the same arguments as in \cite{SS2010}. However, for ease of exposition, here we focus only on the discrete-time case.
\end{rem}

Often, in the system identification framework the distribution of the noise samples is assumed to be Gaussian.
We consider instead the following two models for the noise:
\begin{itemize}
	\item the Laplacian distribution, where the probability density function (pdf) is given by
	\begin{equation}
	p(v_t) = \frac{1}{\sqrt 2 \sigma} e^{- \frac{\sqrt 2 |v_t|}{\sigma}} \,;
	\end{equation}
    %with $\mathbb E [v_t^2] = \sigma^2$;
	\item the Student's t distribution, where the pdf is given by
	\begin{equation} \label{eq:students}
	\!\!\!\!\!\!\!\!\!p(v_t|\nu) = \frac{\Gamma\left(\frac{\nu+1}{2}\right)}{\Gamma\left(\frac{\nu}{2}\right)\sqrt{\pi (\nu-2)\sigma^2}} \left(1+ \frac{v_t^2}{(\nu-2)\sigma^2}\right)^{-\frac{\nu+1}{2}}.
	\end{equation}
    The above equation represents a family of densities parameterized by $\nu$, which is usually called \emph{degrees of freedom} of the distribution.  We will discuss choice of this parameter in Section \ref{sec:estimating_nu}.
\end{itemize}
In both cases we have $E [v_t^2] = \sigma^2$; note that the variance of the Student's t-distribution is defined only for $\nu >2 $. Independently of the model employed in our identification scheme, we shall assume that the noise variance $\sigma^2$ has been consistently estimated by first fitting a long FIR model to the data  (using standard least-squares) and then computing the sample variance of the residuals.
%
%using the consistent estimator defined by the following steps:
%\begin{enumerate}
%\item compute the least-squares estimate of $g$, i.e.
%\begin{equation}
%\hat g_{LS} = (U^T U)^{-1}U^T y \,,
%\end{equation}
%in order to obtain an unbiased estimate of $g$;
%\item compute the empirical estimate of $\sigma^2$
%\begin{equation} \label{eq:hat_sigma}
%\hat \sigma^2 =  \frac{\left(y-U\hat g_{LS}  \right)^{T}\left(y-U\hat g_{LS} \right)}{N-n} \,.
%\end{equation}
%\end{enumerate}
%In particular, when the Student's t is adopted as model, $\rho$ can be retrieved through the relation
%\begin{equation} \label{eq:var_Student}
%\rho = \frac{\nu-2}{\nu} \sigma^2 \,.
%\end{equation}

%we shall consider $\nu$ as an user parameter, whereas
%%Giulio, it is possible to also estimate the nu from data using an ML approach; you just don't get a closed form formula I think.
% But I don't think this is an important point.

\section{Bayesian modeling of system and noise} \label{sec:bayesian}
In this section we describe the probabilistic models adopted for the quantities of interest in the problem.
\subsection{The stable spline kernel and system identification under Gaussian noise assumptions} \label{sec:ssk}
We hereby briefly review the standard kernel-based system identification technique under Gaussian noise assumptions. We first focus on setting a prior on $g$. Following a Gaussian process regression approach \cite{Rasmussen}, we model $g$ as a zero-mean Gaussian random vector, i.e.
\begin{equation}\label{eq:model_g}
p(g) \sim \mathcal N (0,\,\lambda K_{\beta}) \,,
%g \sim \mathcal N (0,\,\lambda K_\beta) \quad,\, \lambda \geq 0 \,.
\end{equation}
where $K_\beta$ is a covariance matrix whose structure depends on  the parameter $\beta$, and $\lambda \geq 0$ is a scaling factor.
%determining the amplitude of the realizations of $g$.
In this context, $K_\beta$ is usually called a {\it kernel} (due to the connection between Gaussian process regression and the theory of reproducing kernel Hilbert space, see e.g. \cite{Wahba1990} and \cite{Rasmussen} for details) and determines the properties of the realizations of $g$.
%whose variance may depend upon few . More specifically, we assume $g$ as
%In the above, the covariance matrix is proportional  to a matrix $K_\beta$ (which depends on the hyperparameter $\beta$ described in the following), which is usually called kernel and determines the %properties of the realizations of $g$.
In this paper, we choose $K_\beta$ from the class of the stable spline kernels \cite{SS2010}, \cite{SS2011}. In particular we shall make use of the so-called \emph{first-order stable spline kernel} (or \emph{TC kernel} in \cite{ChenOL12}), defined as
\begin{equation} \label{eq:ssk1}
\{K_\beta\}_{i,j} := \beta^{ \max(i,j)} \,.
\end{equation}
In the above equation, $\beta$ is a scalar in the interval $[0,\,1)$ and regulates the velocity of the decay of the generated impulse responses.
%Another possible choice of kernel (which we shall not explore in this paper) is the \emph{second-order stable spline kernel}, defined by
%\begin{equation}\label{eq:ssk2}
%\{K_\beta\}_{i,j} = \left[ \frac{\beta^{ (i+j)}
%\beta^{\max(i,j)}}{2}-\frac{\beta^{3 \max(i,j)}}{6}\right] \,.
%\end{equation}
%Compared to \eqref{eq:ssk1}, the latter type of stable spline kernel generates smoother impulse responses. Other kernels for linear system identification are found in \cite{ChenOL12}, \cite{chen2014constructive}.
%%Giulio, why provide the formula for a kernel you are not using in the paper? :) Though I must admit I liked seeing it.

Let us assume that the noise $v$ is Gaussian (see e.g. \cite{SS2010}, \cite{ChenOL12}, \cite{pillonetto2014kernel}).
In this case,  the joint distribution of the vectors $y$ and $g$, given values of $\lambda$ and $\beta$, is jointly Gaussian.
%, namely
%\begin{equation} \label{eq:joint_Gaussian}
%p\left(\left.\begin{bmatrix} y \\ g \end{bmatrix}\right|\lambda,\,\beta \right) \sim \mathcal N \left( \begin{bmatrix} 0\\0 \end{bmatrix} , \begin{bmatrix} \Sigma_y & \Sigma_{yg} \\ \Sigma_{gy} & \lambda K_\beta \end{bmatrix} \right)\,,
%\end{equation}
%where $\Sigma_{yg} = \Sigma_{gy}^T =  \lambda U K_\beta$, and
%\begin{equation} \label{eq:var_y}
%\Sigma_y = \lambda U K_\beta U^T + \Sigma_v \,.
%\end{equation}
It follows that the posterior of $g$, given $y$ (and values of $\lambda$ and $\beta$) is Gaussian, namely
%\begin{equation} \label{eq:py}
%p(y|g,\,\theta) = \mathcal N\left(Ug,\, \Sigma_v \right)
%\end{equation}
%and
\begin{equation} \label{eq:pg}
p(g|y,\,\lambda,\,\beta) = \mathcal N\left(Cy,\,P \right) \,,
\end{equation}
where
\begin{align} \label{eq:CandP}
P & = \left( U^T\Sigma_v^{-1}U + (\lambda K_\beta)^{-1} \right)^{-1}  \\
C & = P U^T \Sigma_v ^{-1} \nonumber \,.
\end{align}
In~\eqref{eq:CandP}, $\Sigma_v$ represents the covariance matrix of the noise vector $v$; in this case  we have $\Sigma_v = \sigma^2 I_N$. Equation \eqref{eq:pg} is instrumental to derive our impulse response estimator, which can be obtained as its minimum mean square error (MSE) (or Bayesian) estimate \cite{Anderson:1979}
\begin{equation} \label{eq:Bayesest_gaussian}
\hat g = \mathbb E [g|y,\,\lambda,\,\beta] = C y \,.
\end{equation}

%
%
%The  estimation of $g$ is given by its Bayesian linear estimate , namely
%\begin{equation} \label{eq:bayes_est1}
%\hat g  = \mathbb E[g|y,\,\lambda,\,\beta] = \Sigma_{gy}\Sigma_y^{-1}y \,
%\end{equation}
The above equation depends on  hyperparameters $\lambda$ and $\beta$.
Estimates of these parameters, denoted by $\hat \lambda$ and $\hat \beta$,
can be obtained by exploiting the Bayesian problem formulation.
More precisely, since $y$ and $g$ are jointly Gaussian, an efficient method to choose $\hat \lambda$ and $\hat \beta$ is given by maximization of the marginal likelihood \cite{pillonetto2014tuning}, which is obtained by integrating out $g$ from the joint probability density of $(y,\,g)$.  Then we have
\begin{align}\label{eq:marglik}
(\hat \lambda, \hat \beta) & = \arg \max_{\lambda,\beta} p(y|\lambda,\,\beta) \nonumber\\
			& =\arg \min_{\lambda,\beta} \log \det (\Sigma_y) + y^T \Sigma_y^{-1} y \,,
\end{align}
where $\Sigma_y = \lambda U K_\beta U^T + \Sigma_v$ is the variance of the vector $y$.
\subsection{Modeling noise as a scale mixture of Gaussians} \label{sec:mixture}

The assumptions on the noise distribution adopted in this paper imply that the joint probabilistic description of $g$ and $y$ is not Gaussian,
and so the method briefly described in the previous section does not apply.
In this section, we show how to deal with this problem.
The key idea is to represent the noise samples $\{v_t\}_{t=1}^N$ as a scale mixture of normals \cite{andrews1974}.
Specifically, with the noise models adopted in this paper, the pdf of each variable $v_t$ can always be expressed as
\begin{equation} \label{eq:scale_mixture}
p(v_t) =  \int_0^{+\infty}  \frac{1}{\sqrt{2 \pi \tau_t}} e^{- \frac{v_t^2}{2 \tau_t}} p(\tau_t) d \tau_t \,,
\end{equation}
where $p(\tau_t)$  is a proper pdf for $\tau_t$.
Since
\begin{equation} \label{eq:noise_prob}
p(v_t) =  \int_0^{+\infty}  p(v_t,\,\tau_t) d \tau_t =  \int_0^{+\infty}  p(v_t| \tau_t) p(\tau_t) d \tau_t \,,
\end{equation}
by comparing \eqref{eq:noise_prob} with \eqref{eq:scale_mixture} we have
\begin{equation}
p(v_t|\tau_t) = \frac{1}{\sqrt{2 \pi \tau_t}} e^{- \frac{v_t^2}{2 \tau_t}} \,,
\end{equation}
which is a Gaussian random variable. Hence, each sample $v_t$ can be thought of as generated in two steps.
\begin{enumerate}
\item A random variable $\tau_t$ is drawn from the pdf $p(\tau_t)$
\item $v_t$ is drawn from a Gaussian distribution with variance $\tau_t$.
\end{enumerate}
%The above expression highlights the fact that each noise sample can be thought of as a realization of a Gaussian random variable,
The distribution of $\tau_t$ depends on the model of $v_t$.
\begin{enumerate}
\item When $v_t$ is Laplacian, we have
    \begin{equation} \label{eq:p_tau_i_Laplacian}
    p(\tau_t) =  \frac{1}{\sigma^2} e^{- \frac{\tau_t}{\sigma^2}}  \quad, \, \tau_t \geq 0 \,,
    \end{equation}
    i.e., $\tau_t$ is distributed as an exponential random variable with parameter $\sigma^2$.
\item When $v_t$ is modeled using the Student's t-distribution, it follows that
    \begin{equation} \label{eq:p_tau_t_Student}
    p(\tau_t) = \frac{\left(\frac{(\nu-2)\sigma^2}{2}\right)^{\frac{\nu}{2}}}{\Gamma\left(\frac{\nu}{2}\right)} \tau_t^{\left(-\frac{\nu}{2}-1\right)} e^{-\frac{(\nu-2)\sigma^2}{2\tau_t}} \quad, \, \tau_t \geq 0 \,,
    \end{equation}
    which is the probability density of an Inverse Gamma of parameters $\left(\frac{\nu}{2},\,\frac{(\nu-2)\sigma^2}{2}\right)$.
\end{enumerate}
Independently of the noise model, if the value of $\tau_t$ is given, the distribution of the noise samples becomes Gaussian.
%Furthermore, if also the $g$ is given, we have that
%\begin{equation}
%p(y_t|g,\,\tau_t) = \frac{1}{\sqrt{2 \pi \tau_t}} e^{- \frac{(y_t - U_t^T g)^2}{2 \tau_t}} \,,
%\end{equation}
%where $U_t^T$ is the $i$-th row of $U$.

Let
\begin{equation} \label{eq:theta}
\theta := \begin{bmatrix} \lambda & \beta & \tau_1 & \ldots & \tau_N \end{bmatrix} \, \in \mathbb{R}^{N+2} \,.
\end{equation}
Due to the representation of noise as scale mixture of Gaussians, the joint distribution of the vectors $y$ and $g$, given values of $\theta$, is jointly Gaussian. Furthermore, by rewriting the noise covariance matrix as follows
\begin{equation} \label{eq:noise_cov}
\Sigma_v = \diag \{\tau_1,\,\ldots,\,\tau_N\} \,,
\end{equation}
Equations \eqref{eq:pg} and \eqref{eq:CandP} hold, so that our impulse response estimate can still be written
as the Bayesian estimate of $g$, i.e.
\begin{equation} \label{eq:Bayesest}
\hat g = \mathbb E [g|y,\theta] = C y \,.
\end{equation}
Note that this estimator, compared to the estimator in the Gaussian noise case \eqref{eq:Bayesest_gaussian}, depends on the $(N+2)$~-~dimensional vector $\theta$, which we shall call the \emph{hyperparameter vector}.
Hence, our Bayesian system identification algorithm consists of the following steps.
\begin{enumerate}
\item Compute an estimate of $\theta$.
\item Obtain $\hat g$ by computing \eqref{eq:Bayesest}.
\end{enumerate}
In the remainder of the paper, we discuss how to effectively compute the first step of the algorithm.

\section{MAP estimate of hyperparameters via Expectation Maximization} \label{sec:EM}

\subsection{MAP estimation of the hyperparameters}

In the previous section we have seen that the joint description of the output and the impulse response is parameterized by the vector $\theta$. In this section, we discuss how to estimate it from data. First, we give a Bayesian interpretation to the constraints on the kernel hyperparameters. To this end, let us denote by $\chi_{\mathcal{S}}(\cdot)$ the indicator function with support $\mathcal{S}$ and introduce the following notation
\begin{equation}
p(\lambda) \propto \chi_{\mathbb{R}^+}(\lambda) \,,
\end{equation}
which represents a flat (improper) prior accounting for the positivity of the scaling factor $\lambda$. Similarly, we define
\begin{equation}
p(\beta) = \chi_{[0,1)}(\beta) \,,
\end{equation}
according to the constraint $\beta \in [0,\,1)$. Hyperparameters $\lambda,\beta$ are then assumed independent
of each other and of all the variances $\tau_t$.

A natural approach to choose the hyperparameter vector $\theta$ is given by its maximum a posteriori (MAP) estimate, which is obtained by solving
\begin{equation} \label{eq:maxlik}
\hat \theta = \arg \max_{\theta} \log \left(p(y|\theta)p(\theta)\right) \,,
\end{equation}
where $p(\theta)$ is the prior distribution of the hyperparameter vector. In view of the stated assumptions,
recalling also that all the $\tau_t$ are independent identically distributed, we have
\begin{equation} \label{eq:prior_theta}
p(\theta) = p(\lambda) p(\beta) \prod_{t=1}^N p(\tau_t).
\end{equation}
%since the $\tau_t$ are independent identically distributed and independent of $\lambda$ and $\beta$ (which are also %independent of each other).
\subsection{The EM scheme} \label{sec:EM_scheme}
Solving \eqref{eq:maxlik} in that form can be hard, because it is a nonlinear and non-convex problem involving $N+2$ decision variables. For this reason, we propose an iterative solution scheme using the EM method.
To this end, we introduce the \emph{complete likelihood} $\log (p(y,g|\theta)p(\theta))$;
the solution of \eqref{eq:maxlik} is obtained by iteratively marginalizing over $g$, which plays the role of \emph{missing data} or \emph{latent variable}. Each iteration of the EM scheme provides an estimate of the hyperparameter vector, which we denote by $\hat \theta^{(k)}$ (where $k$ indicates the iteration index).
Let us define
\begin{equation}
L(y,g|\theta)  := \log p(y,g|\theta) \,,
\end{equation}
where $g$ here is  the latent (unavailable) variable. For ease of notation, we also define
\begin{equation}
L(\tau_t)  :=  \log p(\tau_t)     \,.
\end{equation}
%For convenience, below we shall write $L(\tau_1,\,\ldots,\,\tau_N)$ to indicate the non-dependence of the prior on $\lambda$ and $\beta$.
Then, the EM method provides $\hat \theta$ by iterating the following steps:
\begin{description}
\item[(E-step)] Given an estimate $\hat \theta^{(k)}$, compute
    \begin{equation}
    Q(\theta,\,\hat \theta^{(k)}) := \mathbb{E}_{p(g|y,\,\hat \theta^{(k)})} \left[L(y,g|\theta) + \sum_{t=1}^N  L(\tau_t) \right] \,;
    \end{equation}
\item[(M-step)] Compute
    \begin{equation}
    \hat \theta^{(k+1)} = \arg \max_{\theta}  Q(\theta,\,\hat \theta^{(k)}) \,.
    \end{equation}
\end{description}

 The main advantage of employing the EM method is that convergence to a (local or global) maximum of the objective function is guaranteed \cite{mclachlan2007EM}, \cite{tseng2004analysis}.

\subsection{A posteriori total residual and differential impulse response energy}

As will be seen in the following subsection, our EM scheme for robust system identification procedure
relies on two key quantities which will be defined below and called {\it{a posteriori total residual energy}}
and {\it{a posteriori differential impulse response energy}}.\\
Assume that, at  iteration $k+1$ of the EM scheme, the estimate $\hat \theta^{(k)}$ of $\theta$ is available.
Using the current estimate of the hyperparameter vector, we construct the matrices $\hat C^{(k)}$ and $\hat P^{(k)}$ using \eqref{eq:CandP} and, accordingly, we denote by $\hat g^{(k)}$ the estimate of $g$ computed using \eqref{eq:Bayesest}, i.e. $\hat g^{(k)} = \hat C^{(k)}y$. The linear predictor of $y$ is
\begin{equation} \label{eq:predictor}
\hat y^{(k)} = U \hat g^{(k)} \,,
\end{equation}
with covariance matrix
\begin{equation} \label{eq:pred_variance}
\hat S^{(k)} = U \hat P^{(k)} U^T \,.
\end{equation}
Then, we define the {\it{a posteriori total residual energy}} at time instant $t$ as
\begin{equation}\label{RE}
\hat \varepsilon_t^{(k)} := (y_t - \hat y_t^{(k)})^2 + \hat s_{tt}^{(k)}  \,,
\end{equation}
where $\hat s_{tt}^{(k)}$ is the $t$-th diagonal element of $\hat S^{(k)}$.
Note that $\hat \varepsilon_t^{(k)}$
is the sum of a component related to the adherence to data and a term accounting
for the model uncertainty (represented by $\hat S^{(k)}$).\\
Now, let
\begin{equation} \label{eq:matrix_T}
\Delta := \begin{bmatrix}
			1	& -1			&			&	0\\	
					&	1		&		\ddots	&	\\	
				& 	& \ddots	&-1	\\	
			0	&			& 		& 1	
		\end{bmatrix} \,,
\end{equation}
and define
\begin{equation} \label{eq:transformation1}
\widehat {\delta g}^{(k)}:= \Delta \hat g^{(k)}
\end{equation}
and
\begin{equation} \label{eq:transformation2}
\hat H^{(k)} := \Delta \hat P^{(k)} \Delta^{T} \,.
\end{equation}
Note that $\Delta$ acts as a discrete derivator, so that $\widehat {\delta g}^{(k)}$ represents the estimate of the discrete derivative of the impulse response $g$ (using the current hyperparameter vector $\hat \theta^{(k)}$). The matrix $\hat H^{(k)}$ is its a posteriori covariance. Then, denoting by $\hat h^{(k)}_{ii}$ the $i$-th diagonal element of $\hat H^{(k)}$,
we define the {\it{a posteriori differential impulse response energy}} at $i$ as
\begin{equation}\label{DIRE}
\hat d_i^{(k)}  :=  \left(\widehat {\delta g}_i^{(k)}\right)^2 + \hat h^{(k)}_{ii}.
\end{equation}
Note that this quantity is the sum of the energy of the estimated impulse response derivative
and a term accounting for its uncertainty.

\subsection{Robust EM kernel-based system identification procedure}

The following theorem states how to solve \eqref{eq:maxlik} using the EM method.

\begin{thm} \label{th:min_tau}
Let $\hat \theta^{(k)}$ be the estimate  of the hyperparameter vector at the $k$-th iteration of the EM method, employed to solve \eqref{eq:maxlik}. Then, the estimate $\hat \theta^{(k+1)}$ is obtained with the following update rules:
\begin{itemize}
\item Depending on the noise model, for any $\tau_t,\,t=1,\,\ldots,\,N$ we have
	\begin{enumerate}
	\item In the case of Laplacian distribution,
    	\begin{equation} \label{eq:min_Laplacian}
    	\hat \tau_t^{(k+1)} = \frac{\sigma^2}{4} \left(\sqrt{ 1 + \frac{8\hat \varepsilon_t^{(k)}}{\sigma^2}} - 1 \right) \,;
    	\end{equation}
	\item  In the case of Student's t-distribution,
    	\begin{equation} \label{eq:min_Student}
    	\hat \tau_t^{(k+1)} = \frac{\hat \varepsilon_t^{(k)} + (\nu-2)\sigma^2}{\nu+3} \,;
    	\end{equation}
	\end{enumerate}
%\item The hyperparameter $\beta$ is obtained solving
%	\begin{equation} \label{eq:min_beta}
%	\hat \beta^{(k+1)} = \arg \min_{\beta \in (0,1)} Q(\beta) \,,
%	\end{equation}
%	where
%	\begin{equation} \label{eq:Q_beta}
%	 Q(\beta)\! := \! n \log \Tr \left\{K_\beta^{-1} (\hat g^{(k)} \hat g^{(k)T} + \hat P^{(k)}) \right\} + \log \det K_\beta
%	 \end{equation}
%\item The hyperparameter $\lambda$ is obtained computing
%	\begin{equation} \label{eq:min_lambda}
%	\hat \lambda^{(k+1)} = \frac{1}{n} \Tr \left\{K_{\hat \beta^{(k+1)}}^{-1} (\hat g^{(k)} \hat g^{(k)T} + \hat P^{(k)}) \right\} \,,
%	\end{equation}
%	where $\beta^{(k+1)}$ is the solution of \eqref{eq:min_beta}.
\item The hyperparameter $\beta$ is obtained solving
	\begin{equation} \label{eq:min_beta}
	\hat \beta^{(k+1)} = \arg \min_{\beta \in [0,1)} Q(\beta) \,,
	\end{equation}	
	where	
	\begin{equation} \label{eq:Q_beta}
	Q(\beta)  := n \log f(\beta) +  \frac{n(n+1)}{2} \log \beta + (n-1) \log(1-\beta) \,,
	\end{equation}
	and
	\begin{equation} \label{eq:function_f}
	f(\beta) := \sum_{i=1}^{n-1} \hat d_i^{(k)}  \beta^{1-i} + \hat d_n^{(k)}  		(1-\beta)\beta^{1-n} \,;
	\end{equation}
\item  The hyperparameter $\lambda$ is obtained computing
	\begin{equation} \label{eq:min_lambda}
	\hat \lambda^{(k+1)} = \frac{1}{n} \sum_{i=1}^n  \hat d_i^{(k)} w_{\hat \beta^{(k+1)},i}
	\end{equation}
	where
	\begin{equation} \label{eq:vector_w}
	w_\beta  := \frac{1}{1 - \beta} \begin{bmatrix} \beta^{-1} & \ldots & \beta^{1-n} & (1-\beta) \beta^{-n} \end{bmatrix}
	\end{equation}
	and $w_{\hat \beta^{(k+1)},i}$ are the elements of $w_\beta$ when $\beta = \hat \beta^{(k+1)}$.
\end{itemize}
\end{thm}

The result of the above theorem is remarkable. It establishes that, employing the EM method to solve \eqref{eq:maxlik}, at each iteration of the EM the estimate of the hyperparameter vector $\theta$ can be obtained by solving a sequence of simple scalar optimization problems. All of them crucially depend on the a posteriori total residual and differential impulse response energy, as defined in (\ref{RE}) and (\ref{DIRE}). Furthermore, the update of any $\tau$ admits a closed-form expression which depends on the adopted noise model. As for the kernel hyperparameters, $\beta$ needs to be updated by solving a problem which, at least in principle, does not admit a closed-form solution. However,
the objective function \eqref{eq:Q_beta} can be evaluated using only computationally inexpensive operations and, since $\beta$ is constrained into the interval $[0,\,1)$, its minimum can be obtained quickly using a grid search.
Once $\beta$ is chosen, the updated value of $\lambda$ is available in closed-form.\\

Algorithm \ref{alg} below summarizes our outlier robust system identification algorithm.
The initial value of the hyperparameter vector $\hat \theta^{(0)}$ can be set to
\begin{equation} \label{eq:theta_0}
\hat \theta^{(0)} = \begin{bmatrix} \hat \lambda_{ML} &\hat \beta_{ML}& \sigma^2& \ldots& \sigma^2 \end{bmatrix} \,,
\end{equation}
where $\hat \lambda_{ML}$ and $\hat \beta_{ML}$ are obtained from \eqref{eq:marglik}, i.e. by maximizing the marginal likelihood related to the Gaussian noise case, while the choice $\sigma^2$ is motivated by the fact that $\mathbb{E}[\tau_t] = \sigma^2$ in both the Laplacian and Student's t noise models\footnote{The expectation here is w.r.t. pdf of the $\tau_t$.}. In our experimental analysis (see Section \ref{sec:numerical}), this choice of initialization has always led the EM method to attain the global maximum of the objective function.

%\noindent\rule{\columnwidth}{0.3mm} \vspace{-.5cm}
\begin{alg}\label{alg}\emph{\textbf{: EM-based outlier robust system identification} \vspace{0.1cm}\\
Input: $\{y(t)\}_{t=1}^N,\,\{u(t)\}_{t=0}^{N-1}$ \vspace{0.1cm} \\
Output: $\{\hat{g_t}\}_{t=1}^n$
\begin{enumerate}
\item Initialization: Set $\hat \theta^{(0)}$;
\item Repeat until $\frac{\|\hat \theta^{(k+1)} - \hat \theta^{(k)}\|}{\|\hat \theta^{(k)}\|}$ is below a given 			threshold or a prescribed number of iterations is reached:
    \begin{enumerate}
    \item  Update the a posteriori total residual $\hat \varepsilon_t^{(k)}$ and the differential impulse response energy
    $\hat d_i^{(k)} $ according to (\ref{RE}) and (\ref{DIRE}), respectively;
        \item  Update $\hat \tau_t^{(k+1)}$ using \eqref{eq:min_Laplacian} or \eqref{eq:min_Student};
        \item Update $\hat \beta^{(k+1)}$ solving \eqref{eq:min_beta};
        \item Update $\hat \lambda^{(k+1)}$ computing \eqref{eq:min_lambda};
	\end{enumerate}
\item Compute $\hat g$ as in \eqref{eq:Bayesest}.
\end{enumerate}}
\end{alg}

%\vspace{-.6cm}
%\noindent\rule{\columnwidth}{0.3mm}
\begin{rem}
Note that the iteration complexity depends also on the number of operations required
to update the a posteriori total residual $\hat \varepsilon_t^{(k)}$ and the differential impulse response energy
$\hat d_i^{(k)}$. In view of the definitions (\ref{RE}) and (\ref{DIRE}), this is related to
the computation of the posterior covariance $\hat P^{(k)}$ which,
recalling (\ref{eq:CandP}), requires the inversion of an $n \times n$ matrix.
Since $n$ is the number of unknown impulse response coefficients,
this result is also computational appealing since, in system identification,
typically one has $n \ll N$, where $N$ is the data set size.

%while $N$ is the data set size.
%Then, at every step, the subproblems are $N+2$, defined
%just inverting a matrix of size $n \times n$ (typically, $n \ll N$ in system identification).
\end{rem}

\begin{rem}
Regarding the noise model induced by the Student's t distribution, when the parameter $\nu$ is set to 2, the prior \eqref{eq:p_tau_t_Student} becomes flat, i.e. $p(\tau_t) \propto \chi_+(\tau_t),\, t = 1,\,\dots,\,N$. Hence, the update rule \eqref{eq:min_Student} becomes
\begin{equation}
\hat \tau_t^{(k+1)} = \hat \varepsilon^{(k)} \,,
\end{equation}
i.e., no information on the noise variance is used and the values of the $\tau_t$ are completely estimated from the data. This is in accordance with the fact that the Student's t-distribution has infinite variance for $\nu=2$, which means that the prior on noise is not carrying information (from the second order moments point of view). Conversely, when $\nu = +\infty$ Equation \eqref{eq:min_Student} becomes
\begin{equation}
\hat \tau_t^{(k+1)} = \sigma^2 \,,
\end{equation}
which means that all the noise samples must have the same variance, equaling $\sigma^2$. This reflects the fact that a Student's t-distribution with $\nu = +\infty$ is in fact a Gaussian distribution (in this case with variance equal to $\sigma^2$), so that no outliers are expected by this noise model.
\end{rem}

\section{Extensions of the algorithm} \label{sec:extensions}

\subsection{Estimating the degrees of freedom of the Student's t-distribution} \label{sec:estimating_nu}

%Since Fig. \ref{fig:results_WN} and \ref{fig:results_LP} show that the parameter $\nu$ of the Student's t distribution influences the performance of the estimator, we have also investigated determination of $\nu$ from data.
%A possible way is to still use the MAP estimator \eqref{eq:maxlik}: the estimate becomes
%\begin{equation} \label{eq:nu_est}
%\hat \nu = \arg \max_{\nu} \log (p(y|\hat \theta)p(\hat\theta)) \,,
%\end{equation}
%where $\hat \theta$ is the estimated value of the hyperparameter vector.\\
%This approach has been applied to the same data generated for the second experiment (filtered white noise as input).
%The set of candidate values for $\nu$ is the same as above, i.e. $\{2,\,2.25,\,2.5,\,2.75,\,3,\,5\}$. The value chosen by the
%estimator \eqref{eq:nu_est} is always $\hat \nu = 2.25$. Table 1 %\ref{tab:nuest}
%shows a comparison between the
%estimator EM-S-$\nu$, with $\nu$ chosen by \eqref{eq:nu_est} (we call this estimator EM-S-$\hat \nu$) and the estimator
%EM-S-opt, which selects the value of $\nu$ maximizing the fitting score (\ref{Fiti}).
%The performance of the two estimators is very similar.

The parameter $\nu$ of the Student's t-distribution \eqref{eq:students} affects the algorithm capability of detecting outliers. It is thus desirable to have an automatic procedure for tuning this parameter. In this section we show how to include the estimation of $\nu$ in the proposed EM scheme. We treat the estimation of such a parameter as a model selection problem \cite{Chipman01thepractical}, where we aim at choosing $\nu$ maximizing the joint distribution of $y$ and $g$, obtained by integrating out the $\tau_t$. More precisely, we have
\begin{align} \label{eq:marg_nu}
p(y,\,g|\nu) &= \int p(y,\,g|\tau_1,\,\ldots,\,\tau_N,\,\nu) \prod_{t=1}^N p(\tau_t|\nu) \prod_{t=1}^N d \tau_t \nonumber \\
			 & \propto \int p(y|g,\,\tau_1,\,\ldots,\,\tau_N,\,\nu) \prod_{t=1}^N p(\tau_t|\nu) \prod_{t=1}^N d \tau_t \nonumber \\
			 &= p(v|\nu) \,,
\end{align}
where the second step follows from the fact that $g$ is independent of the $\tau_t$ and thus the terms $p(g|\tau_t)$ can be neglected.
We now suppose that also $\nu$ is included in the iterative scheme and we assume that at the $k$-th iteration we have obtained the estimate $\hat \nu^{(k)}$.
At each EM iteration, we can compute an estimate of the noise samples given by
\begin{equation} \label{eq:noise_est}
\hat v^{(k)} = y - \hat y^{(k)}\,,
\end{equation}
where $\hat y^{(k)}$ is the linear predictor introduced in \eqref{eq:predictor}. Since   each sample $v_t$ follows a Student's t-distribution (and is independent of $v_j$, $j\neq t$), it is natural to choose the value of $\nu$ maximizing the log-likelihood
\begin{align} \label{eq:likelihood_student}
\hat \nu^{(k+1)} & = \arg \max_{\nu} \, \log \prod_{t=1}^N p(\hat v_t^{(k)}|\nu) \nonumber\\
	& = \arg \max_{\nu} \,  N\log \left( \frac{\Gamma\left(\frac{\nu+1}{2}\right)}{\Gamma\left(\frac{\nu}{2}\right)\sqrt{\pi \sigma^2(\nu-2)}} \right) \\
	& \qquad \qquad\quad- \frac{\nu+1}{2}\sum_{t=1}^N  \log \left( 1 + \frac{(y_t - \hat y_t^{(k)})^2}{\sigma^2(\nu-2)} \right). \nonumber
\end{align}
Although this selection criterion does not rigourously follow the paradigm of the EM method, in the next section we will see that its performance is rather satisfactory.

\subsection{Reducing the number of hyperparameters in the identification process} \label{sec:upsilons}
As seen in Section \ref{sec:bayesian}, the unknown vector $\theta$ contains $N+2$ hyperparameters to be estimated.
If the data set size is large, e.g. $N \sim 10^4$ -- $ 10^5$, even if all of the updates of the $\tau_t$ are decoupled and
consist of one simple operation, it might be desirable to have an even faster identification process. This could be obtained by reducing the size of $\theta$, that is, by constraining groups of $\tau_t$ to assume the same value. Given an integer $p > 0$, let us assume that $m := N/p$ is integer. Then, it is possible to readapt Theorem \ref{th:min_tau} in order to impose the constraints
\begin{align} \label{eq:upsilon_tau}
\Upsilon_1 &:= \tau_1  = \ldots =  \tau_m \,,  \nonumber\\
\Upsilon_2 &:=\tau_{m+1}  = \ldots =  \tau_{2m} \,,  \nonumber\\
&\vdots \\
\Upsilon_p &:=\tau_{(p-1)m+1}  = \ldots =  \tau_{N} \,, \nonumber
\end{align}
so that the new hyperparameter vector
$$\theta = \begin{bmatrix} \lambda & \beta & \Upsilon_1 & \ldots \Upsilon_p \end{bmatrix} \,\, \in \mathbb{R}^{p+2}$$ consists in $p+2$ components (with possibly $p \ll N$).

To this aim, let us introduce the following partition of the matrix $\hat S^{(k)}$ introduced in \eqref{eq:pred_variance}
\begin{equation} \label{eq:partition_S}
\hat S^{(k)} = \begin{bmatrix} \hat S_{11}^{(k)} &\ldots &\hat S_{1p}^{(k)} \\
						\vdots &\ddots & \vdots \\
						\hat S_{p1}^{(k)} &\ldots &\hat S_{pp}^{(k)} \end{bmatrix} \quad,\, \hat S_{ij}^{(k)} \in \mathbb{R}^{m \times m}
\end{equation}
and, similarly,
\begin{equation} \label{eq:partition_y}
y = \begin{bmatrix}Y_1 \\ \vdots \\ Y_p \end{bmatrix} \quad,\quad \hat y^{(k)} = \begin{bmatrix}\hat Y^{(k)}_1 \\ \vdots \\ \hat Y^{(k)}_p \end{bmatrix} \quad,\, Y_i\,,\,\hat Y^{(k)}_i  \in \mathbb{R}^{m} \,.
\end{equation}
The following result then holds.
\begin{prop}\label{th:upsilon}
Let $\hat \theta^{(k)}$ be the estimate  of the hyperparameter vector at the $k$-th iteration of the EM method. Define
\begin{equation} \label{eq:zeta}
\hat \zeta^{(k)}_i :=  \|Y_i - \hat Y_i^{(k)}\|^2 + \Tr \{\hat S^{(k)}_{ii}\} \quad,\,i=1,\,\ldots,\,p \,.
\end{equation}
Then, depending on the noise model adopted, the estimates $\hat \Upsilon_i^{(k+1)}$, $i=1,\,\ldots,\,p$  are obtained with the following update rules:
\begin{enumerate}
	\item In the case of Laplacian distribution,
    	\begin{equation} \label{eq:min_Laplacian_upsilon}
    	\hat \Upsilon_i^{(k+1)} = \frac{m\sigma^2}{4} \left(\sqrt{ 1 + \frac{8\hat \zeta_i^{(k)}}{m^2\sigma^2}} - 1 \right) \,;
    	\end{equation}
	\item  In the case of Student's t-distribution,
    	\begin{equation} \label{eq:min_Student_upsilon}
    	\hat \Upsilon_i^{(k+1)} = \frac{\hat \zeta_i^{(k)} + (\nu-2)\sigma^2}{\nu+2+m} \,.
    	\end{equation}
\end{enumerate}
\end{prop}
The proof uses arguments similar to those of Theorem \ref{th:min_tau}. Note that the update rules for $\hat \lambda^{(k+1)}$ and $\hat \beta^{(k+1)}$ remain the same as in Theorem \ref{th:min_tau}. The value of the $\Upsilon_i$ can be interpreted as an averaging among the $\tau_t$ ``sharing'' the same $\Upsilon_i$. Clearly, the price to pay for reducing the number of hyperparameters is a lower capability by the algorithm to detect outliers. Note that in principle each $\Upsilon_i$ can be defined so that it corresponds to non-consecutive output measurements. Any choice of groups of $\tau_t$ associated with one $\Upsilon_i$ can be considered, provided that \eqref{eq:partition_S}, \eqref{eq:partition_y} are partitioned accordingly.
\begin{rem} \label{rem:special_case}
In the limit case $p=1$, i.e. $m = N$, it is reasonable to expect that, at least for large indices $k$, $\hat \zeta^{(k)}_i \simeq N\sigma^2$. Then, if $N$ is large, from \eqref{eq:min_Laplacian_upsilon} we get
\begin{equation}
\hat \Upsilon_i^{(k+1)} \simeq \frac{N\sigma^2}{4} \left(\sqrt{ 1 + \frac{8}{N}} - 1 \right) \simeq \frac{N\sigma^2}{4} \frac{4}{N} = \sigma^2 \,,
\end{equation}
whereas \eqref{eq:min_Student_upsilon} can be approximated as
\begin{equation}
\hat \Upsilon_i^{(k+1)} \simeq \frac{N\sigma^2 + (\nu-2)\sigma^2}{\nu+2+N} \simeq \sigma^2 \,.
\end{equation}
Hence, when $p=1$, i.e. when all the $\tau_t$ are forced to converge to the same value, the estimated variance will converge to the nominal noise variance $\sigma^2$ and thus the algorithm will behave as in the Gaussian noise case (see Section \ref{sec:ssk}).
\end{rem}

\section{Experiments} \label{sec:numerical}
In this section we present results of several numerical simulations and describe an experiment on a real system.

\subsection{Monte Carlo studies in presence of outliers} \label{sec:numerical1}

\begin{figure*}[!ht]
\begin{center}
\begin{tabular}{cc}
    \includegraphics[width=7cm]{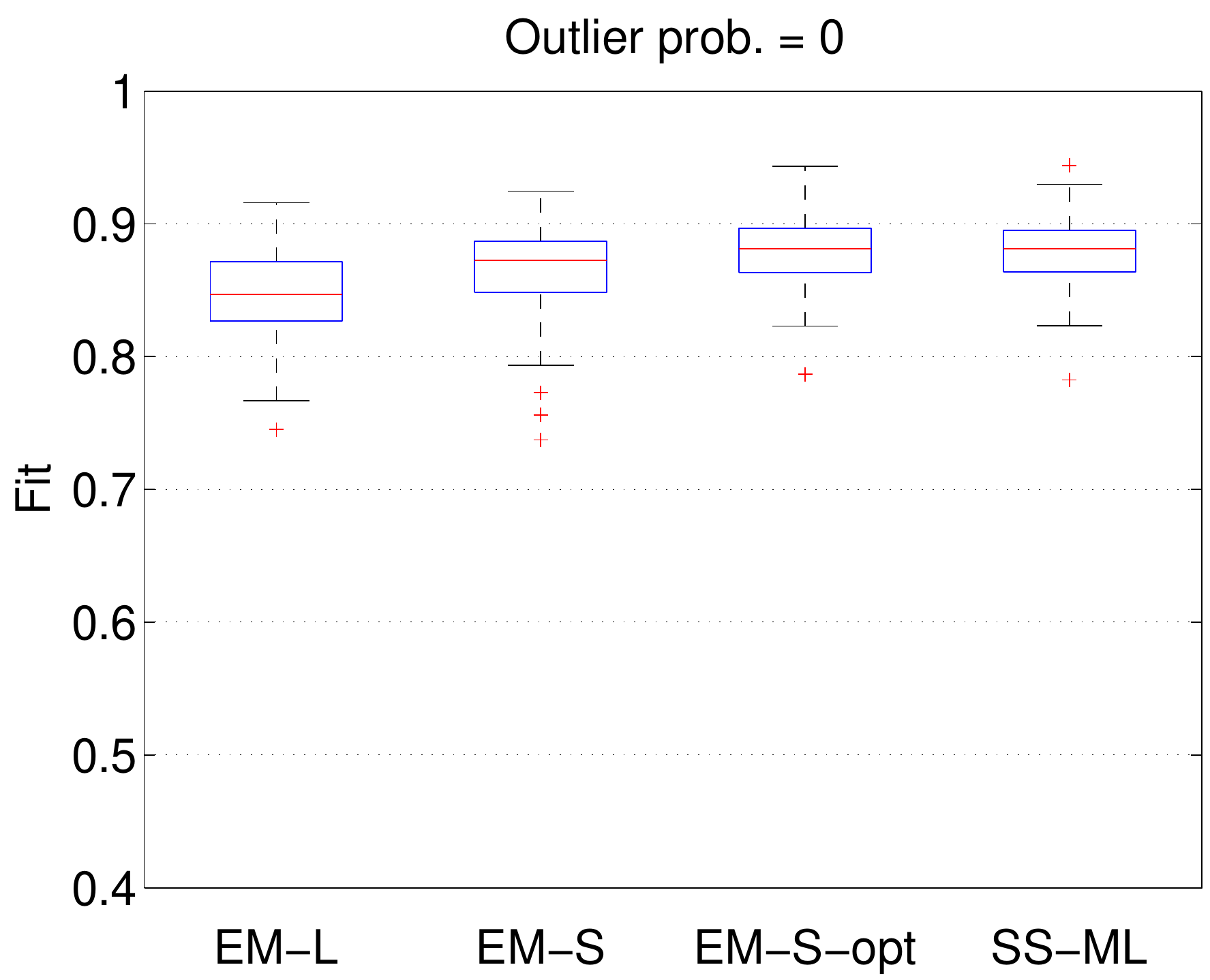} &     \includegraphics[width=7cm]{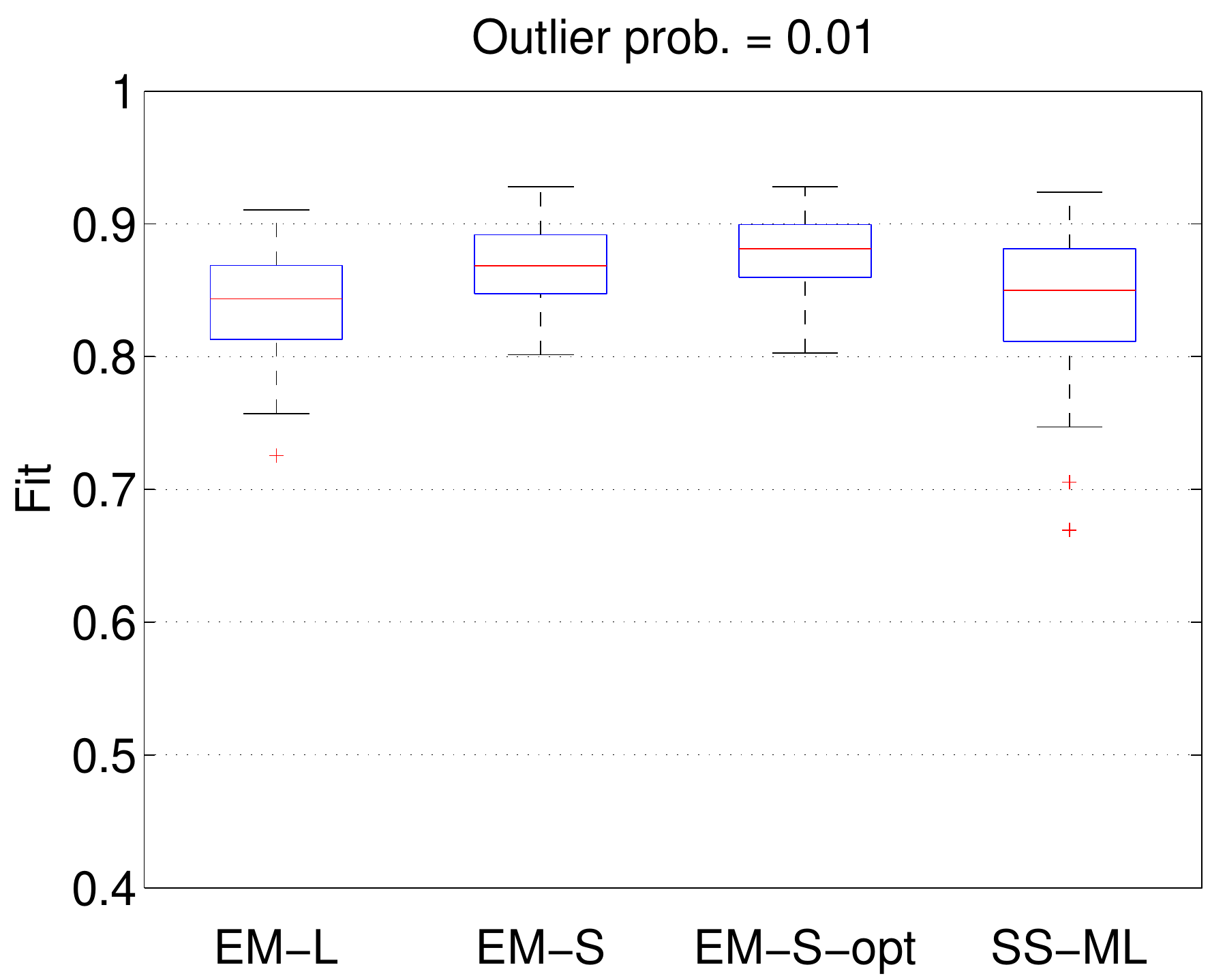} \\
        \includegraphics[width=7cm]{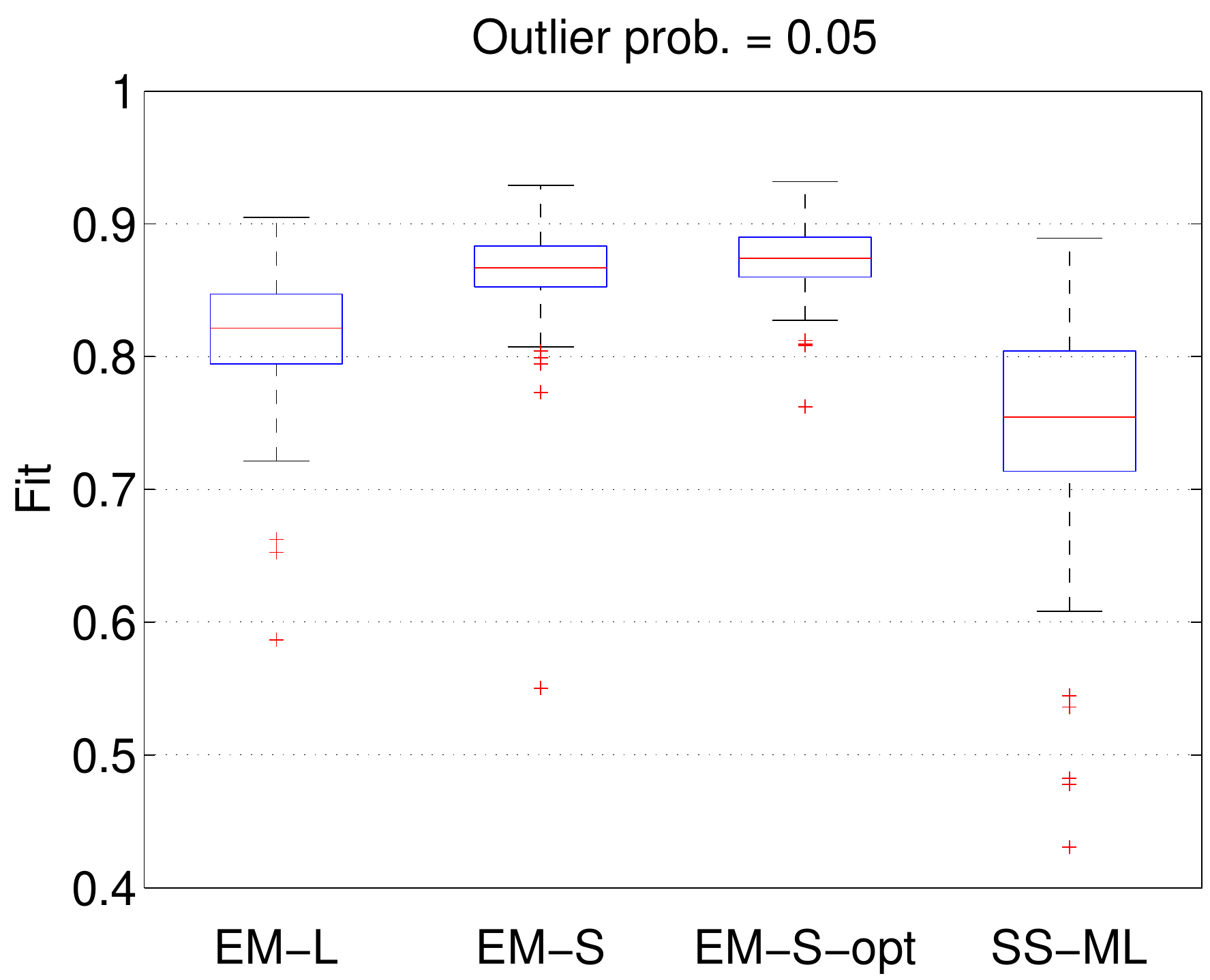} &     \includegraphics[width=7cm]{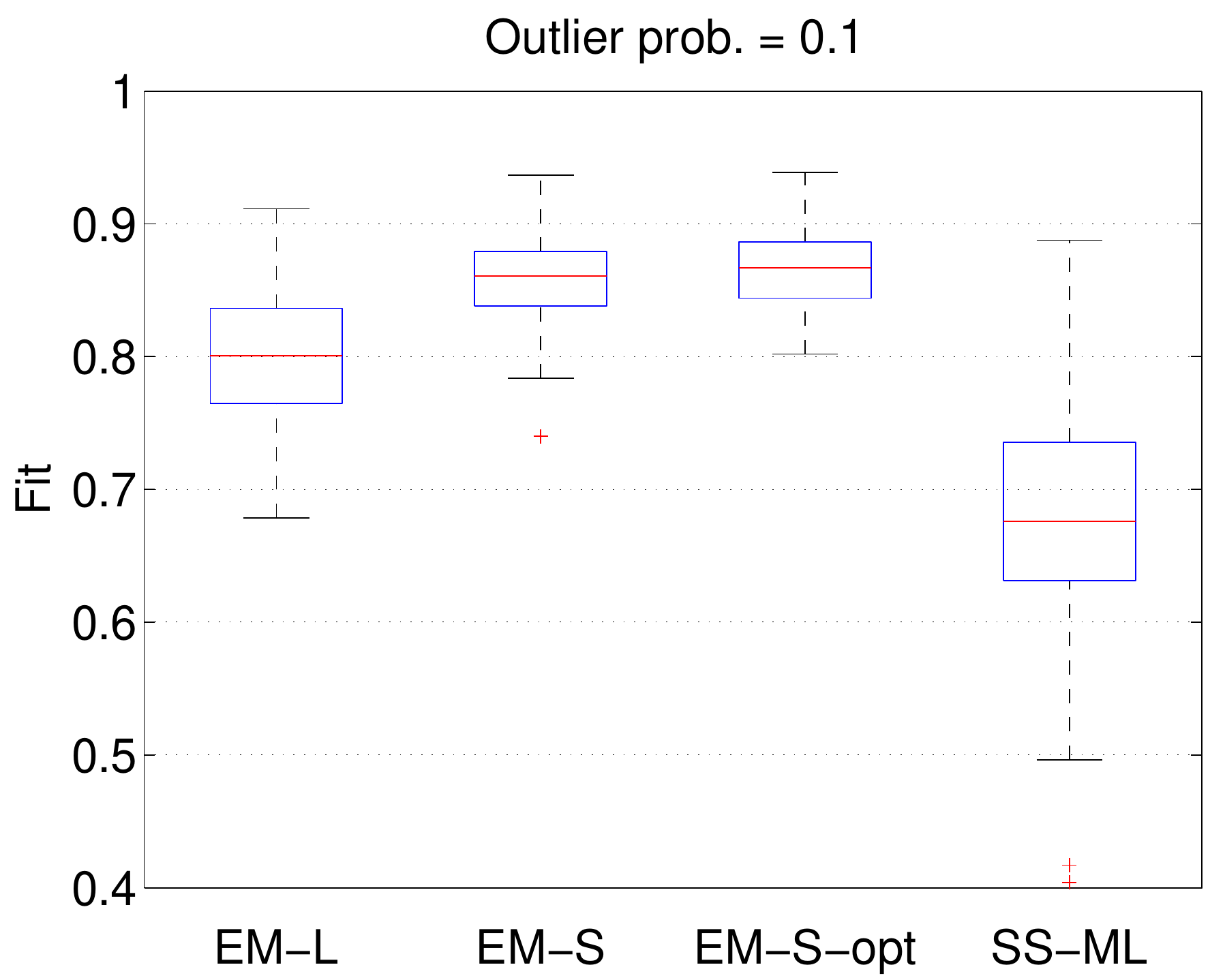}
\end{tabular}
\caption{{\bf{Monte Carlo in presence of outliers (Section \ref{sec:numerical1})}}: Box plots of the fits achieved by the estimators when outliers are present with increasing probability.} \label{fig:results_out}
\end{center}
\end{figure*}
We first perform Monte Carlo simulations to assess the performance of the proposed method. We set up four groups of numerical experiments of 100 independent runs each. At each Monte Carlo run, we generate random dynamic systems of order 30 as detailed in Section 7.2 of \cite{pillonetto2014kernel}. In order to simulate the presence of outliers in the measurement process, the noise samples $v_t$ are drawn from a mixture of
two Gaussians, i.e.
\begin{equation} \label{eq:noise_exp}
v_t \sim (1-c) \mathcal N(0,\sigma^2) + c  \mathcal N(0,100\sigma^2) \,.
\end{equation}
In this way, outliers are measurements with 100 times higher variance,
generated with probability $c$, where $c$ assumes the values $0$, $0.01$, $0.05$ and $0.1$, depending on the experiment. The value of $\sigma^2$ is such that the noise has variance equal to $0.1$ times the variance of the noiseless output. For ease of comparison, the input is always a white noise sequence with unit variance.

Random trajectories of input and noise are generated at each run. The data size is
$N=200$, while the number of samples of the impulse response to be estimated is $n=50$.
The performance of an estimator $\hat g$
is evaluated at any run by computing the fitting score, i.e.
\begin{equation}\label{Fiti}
FIT_i  = 1-\frac{\|g_i - \hat g_i \|_2}{\|g_i\|_2} \,,
\end{equation}
where $g_i$ and $\hat g_i$ represent, respectively, the true
and the estimated impulse responses (truncated at the $n$-th sample) obtained at the $i$-th Monte Carlo run.
In particular, the following estimators $\hat g$ are tested: %methods for estimating $g$.
\begin{enumerate}
\item EM-L: this is the new kernel-based method proposed in this paper adopting a Laplacian model for noise.
Algorithm \ref{alg} is used to estimate hyperparameters. The EM run stops when $\frac{\|\hat \theta^{(k)} - \hat \theta^{(k-1)}\|}{\| \hat \theta^{(k-1)}\|} < 10^{-3}$.
\item EM-S: the same as before except that a Student's t-distribution is used as noise model. The degrees of freedom $\nu$ are chosen according to the procedure described in Section \ref{sec:estimating_nu}, within the grid $\{2.01,\,2.25,\,2.5,\,2.75,\,3,\,5,\,7.5,\, 10,\, 15,\, 50,\, +\infty\}$; the symbol $+\infty$ indicates the Gaussian noise case, to which the Student's t-distribution collapses when $\nu = +\infty$.
\item EM-S-opt: this estimator also makes use of the Student's t-distribution as noise model. The parameter $\nu$ is selected within the grid introduced above by taking the value that maximizes the fit \eqref{Fiti}. To perform this operation, this estimator must have access to the true impulse response and thus it is not implementable in practice.
\item SS-ML: this is the kernel-based identification method proposed in \cite{PillACC2010}, revisited in \cite{ChenOL12} and briefly described in Section \ref{sec:ssk}.
    The impulse response is modeled as in \eqref{eq:model_g} and the hyperparameters $\lambda$ and $\beta$ are estimated by marginal likelihood maximization \eqref{eq:marglik}.
    Note that this estimator does not attempt to model the presence of outliers.
%\item SS-ML-Oracle: this is the same method described above except that
%we choose those values of $\lambda$ and $\beta$ which maximizes the fit (\ref{Fiti}).
%Note that this kind of estimator knows the true impulse response, hence it is not utilizable in practice.
\end{enumerate}

Figure \ref{fig:results_out} shows the results of the simulations. It can be seen that, in general, accounting for outliers pays off in terms of accuracy. Both the proposed robust estimators perform better than the estimator SS-ML, which does not account for the possible presence of outliers. In particular, as the rate of outliers increases, the improvements given by the proposed estimators become more and more significant. Details on the results of the simulations are given in Table 1, which reports the confidence intervals for the average fits of the methods. It is shown that the robust methods give a statistically significant improvements in the performance compared to the estimator SS-ML. A further evidence is given by the one-tailed paired t-test between the estimator SS-ML and the other estimators. For the cases $c = 0.05$, $c = 0.1$ all the three robust estimators significantly outperform SS-ML method (p-value less than $10^{-21}$). For the case $c=0.01$ only EM-S and EM-S-opt significantly outperform SS-ML method (p-value less than $10^{-8}$) while EM-L has comparable performances (p-value equal to $0.797$).
Modeling noise using the Student's t-distribution seems to give better results in terms of fitting. The choice of the degrees of freedoms seems to be very accurate, as the estimator EM-S is very close to EM-S-opt: there is a slight degradation in the performance, which can be motivated as a price to pay for estimating the additional parameter $\nu$.
\begin{table}[h!] \label{tab:fits}
\begin{center}
\begin{tabular}{c|c|c}
 Method 	& $c = 0$ & $c = 0.01$  \\
\hline
 EM-L	 	& $84.63 \pm 0.66$ & $83.92\pm 0.73$  \\
 EM-S       & $86.63 \pm 0.65$ & $86.8\pm 0.61$   \\
 EM-S-opt   & $88.05 \pm 0.5$  & $87.75 \pm 0.55$  \\
 SS-ML      & $87.92 \pm 0.5$  & $84.23 \pm 0.96$ \\
 \hline\hline
Method 	& $c = 0.05$ & $c = 0.1$ \\
\hline
EM-L	 	& $81.63\pm 0.99$	& $79.69\pm 1.03$			\\
EM-S       & $86.11\pm 0.84$	& $85.97\pm 0.64$			\\
 EM-S-opt   & $87.23\pm 0.55$	& $86.54 \pm 0.59$		\\
SS-ML      & $74.65\pm 1.61$ & $67.74 \pm 1.72	$			\\
\end{tabular}
\caption{{\bf{Monte Carlo in presence of outliers (Section {sec:numerical1})}}:
Confidence intervals of the average fits in percent when outliers are present with increasing probability.}
\end{center}
\end{table}
%
%
%\begin{table}
%\begin{center}
%\begin{tabular}{|c|c|c|}
%\hline Method & Median & Mean \\
%\hline
%\hline EM-S-$\hat \nu$ & 0.7819 & 0.7461\\
%\hline EM-S-opt & 0.7831 & 0.7484 \\
%\hline
%\end{tabular}
%\caption{{\bf{Monte Carlo in presence of outliers with degrees of freedom of the Student's t estimated from data (subsection \ref{sec:numerical2}, filtered white noise as input)}}: average fit of the estimator EM-S-$\hat \nu$ selecting $\nu$ by \eqref{eq:nu_est} and of the estimator which has access
%at any run to the optimal value of $\nu$.}
%\end{center}
%\label{tab:nuest}
%\end{table}

\subsection{System identification under Student's t-distributed noise} \label{sec:model_misspecification}
\begin{figure}[!ht]
\begin{center}
    {\includegraphics[width=7cm]{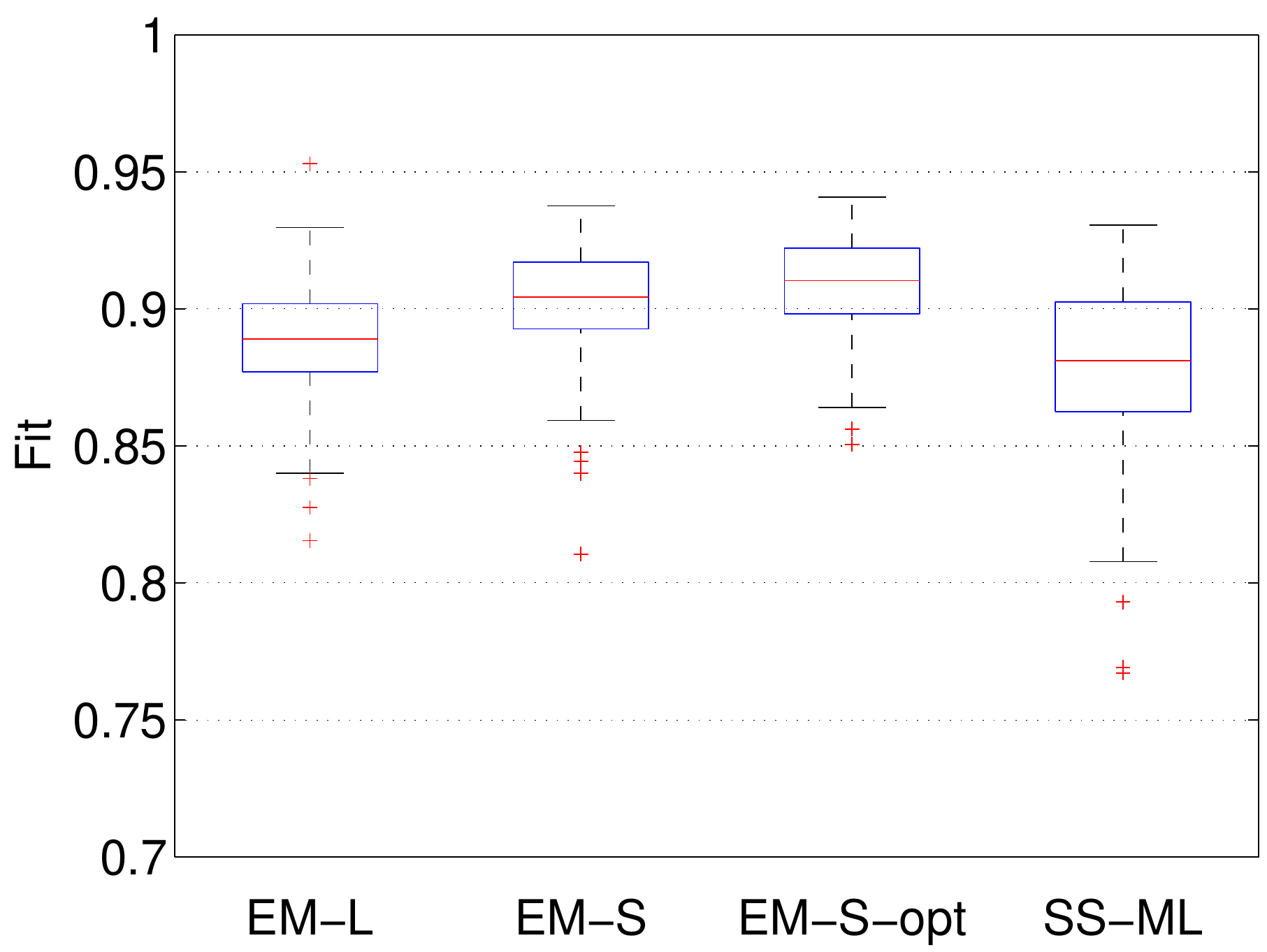}}
\caption{{\bf{Monte Carlo simulations of model misspecification (Section \ref{sec:model_misspecification})}}: Box plots of the fits of the tested estimators.}\label{fig:student_experiment}
\end{center}
\end{figure}

As suggested by an anonymous reviewer, the method proposed in this paper may also by employed in situations where the noise follows the Student's t-distribution. Applications of this scenario are found in model misspecification problems. To this end, we perform an experiment of 100 Monte Carlo runs where the noise samples are drawn from the Student's t-distribution with $\nu=3$ degrees of freedom. The experimental conditions (input and noise variance) are as in the previous experiments. Figure \ref{fig:student_experiment} shows the box plots of the 100 Monte Carlo simulations. The estimator EM-S offers a higher accuracy than the other estimators, since it is able to capture the true noise model. The estimators EM-L and SS-ML offer approximately the same performance. This may be explained by the fact that these two distributions capture different features of the Student's t-distribution: the Laplacian r.v. is heavy-tailed, while the Gaussian r.v. is smooth around zero.
\subsection{Monte Carlo studies in presence of outliers: reduction of the number of hyperparameters} \label{sec:numerical3}

We test  the performance of the estimator described in Section \ref{sec:upsilons},
where the size of the hyperparameter vector is reduced, setting $c=0.3$ and the noise variance equal to 0.01 the noiseless output variance. We compare, by means of 100 Monte Carlo runs, the estimator SS-ML  with a new class of estimators,
dubbed EM-L-$p$. These estimators employ the Laplacian model of noise but, instead of attempting the estimation
of $N$ distinct values of the noise variances $\tau_t$, they estimate $p$ values, as described in Section \ref{sec:upsilons}. In this experiment, we choose 4 different values of $p$, namely $p = 1,\,20,\,40,\,200$, so that the same noise variance $\Upsilon_i,\,i=1,\,\ldots,\,p$, is shared between 200, 10, 5 and 1 output measurements, respectively. Note that the case $p=200$ corresponds to the estimator EM-L described above (since $m=N$), while the case $p=1$ forces all the noise variances to be the same (this thus corresponds to the Gaussian noise case).
\begin{figure}[!ht]
\begin{center}
    {\includegraphics[width=7cm]{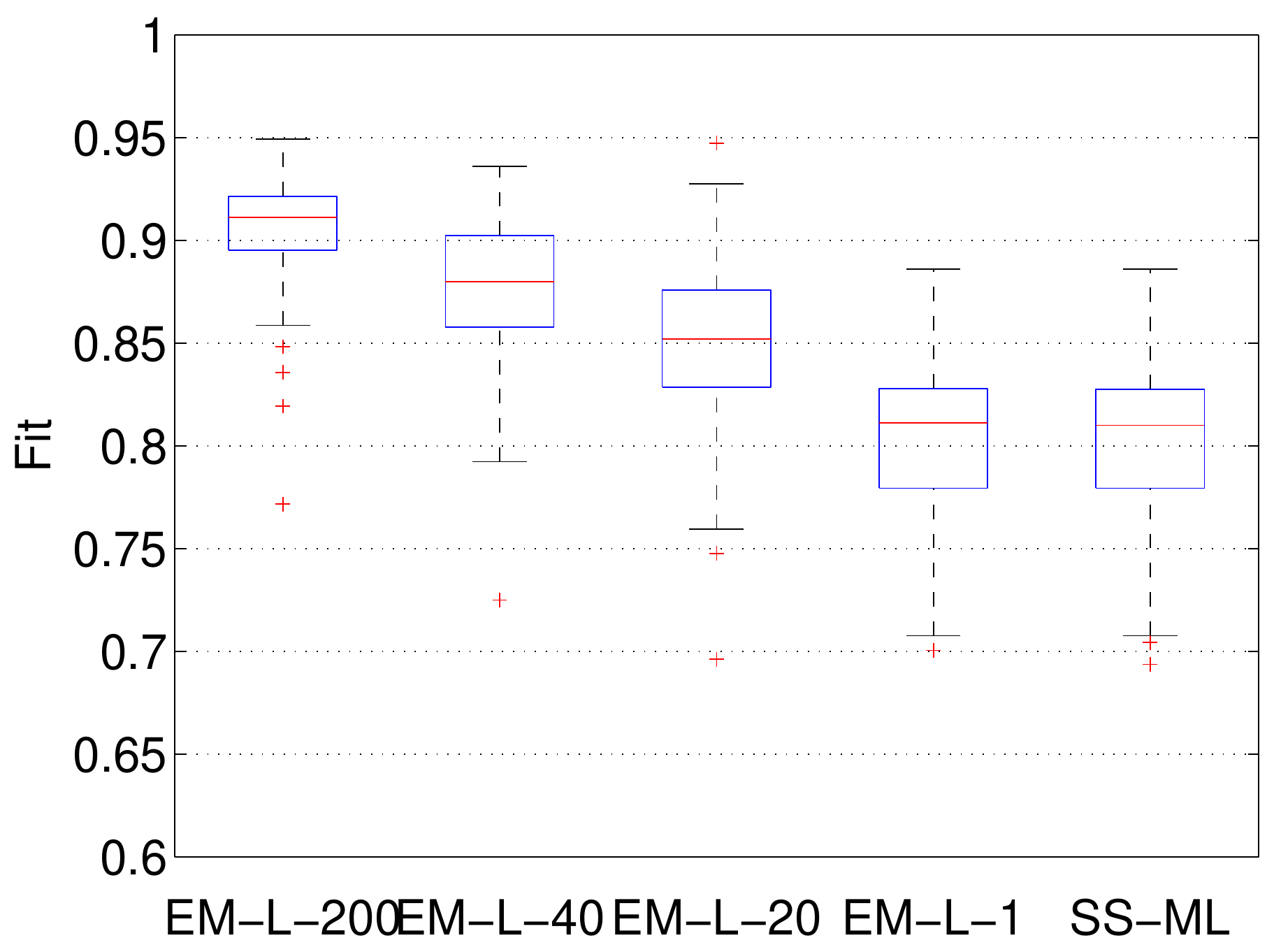}}
\caption{{\bf{Monte Carlo in presence of outliers using a Laplacian noise model with a reduced number of hyperparameters (Section \ref{sec:numerical3})}}: Box plots of the fits with the number $p$ of estimated noise variances ranging from 200 to 1.}\label{fig:upsilon_experiment}
\end{center}
\end{figure}

Figure \ref{fig:upsilon_experiment} shows the result of the experiment. As expected, there is a degradation of the accuracy of the estimators EM-L-$p$ as $p$ decreases: there is a loss of outlier detectability due to the reduction  in the number of noise hyperparameters. However, EM-L-$p$, with $p \neq 1$, still compares favorably with respect to the non-robust estimators. Note that, in accordance with Remark \ref{rem:special_case}, when $p = 1$ the estimator gives almost exactly the same performance of SS-ML (little mismatch is due to different numerical optimization procedures).

%\subsection{Monte Carlo studies in absence of outliers}\label{sec:numerical4}
%
%We also test our method in a case where no outliers are present in the data, i.e. when $c_1 =1$ and $c_2 =0$ in \eqref{eq:noise_exp}. We assume the input to be white noise and we test the same estimators introduced above. From Figure \ref{fig:results_WN_no_outliers}, it can be seen that the performance of the outlier robust methods are slightly worse than the non-robust methods. This is explained by the fact that a non-Gaussian noise model is assumed. However, it can be seen that, in this case, higher values of $\nu$ increase the accuracy of the proposed methods when the Student's t-distribution is adopted as noise model, due to the fact that the Student's t-distribution gets closer and closer to a Gaussian as $\nu$ increases.
%
%\begin{figure*}[!ht]
%\begin{center}
%    {\includegraphics[width=14cm]{N_200_WN_ALL_no_outliers_new.pdf}}
%\caption{{\bf{Monte Carlo in absence of outliers (subsection \ref{sec:numerical4}, white noise as input)}}: box plots of the fits achieved by different estimators.} \label{fig:results_WN_no_outliers}
%\end{center}
%\end{figure*}

\subsection{Comparison with other outlier robust method}\label{sec:comparison_GS}

We compare the performance of the proposed method with the outlier robust system identification algorithm proposed in \cite{bottegal2014outlier}. In this work, the noise is modeled as independent Laplacian r.v.'s; a Gaussian description is obtained via the scale mixture of Gaussians introduced in Section \ref{sec:mixture}. Under a fully Bayesian framework, a Markov Chain Monte Carlo (MCMC) integration scheme based on the Gibbs sampler (see e.g. \cite{Gilks} for details) is used to estimate the impulse response $g$.

We test the method EM-L of Section \ref{sec:numerical1} against the aforementioned algorithm, dubbed GS-L (Gibbs sampler with a Laplacian model of noise), on 100 randomly generated systems under the same conditions as the previous subsection (white noise input, $c = 0.3$, noise variance equal to 0.01 the noiseless output variance). We also evaluate the performance of the method SS-ML. Table 2 reports the median of the fitting scores together with the average computational times required by the two methods. It can be seen that, although the performance of the two robust methods are very close to each other, the computational burden of the method proposed in this paper is much lower. The non-robust method SS-ML is faster, however it returns a lower fit of the impulse responses.
%The method GS-L requires to draw a large number of samples from several conditional distributions,
\begin{table}[h!] \label{tab:comparing_laplacian}
\begin{center}
\begin{tabular}{c|c|c}
 Method & Fit \% (median) & Avg. computational time [s]  \\
\hline
 EM-L & 91.28  & 2.18 \\
 GS-L & 90.78  & 93.77 \\
 SS-ML & 76.59 & 0.18 \\
 \end{tabular}
\caption{{\bf{Comparison of three different identification methods (Section \ref{sec:comparison_GS})}}: Fitting scores and average computational times.}
\end{center}
\end{table}

\subsection{Robust solution of the introductory example}\label{sec:numerical5}

\begin{figure}[!ht]
\begin{center}
    {\includegraphics[width=8.5cm]{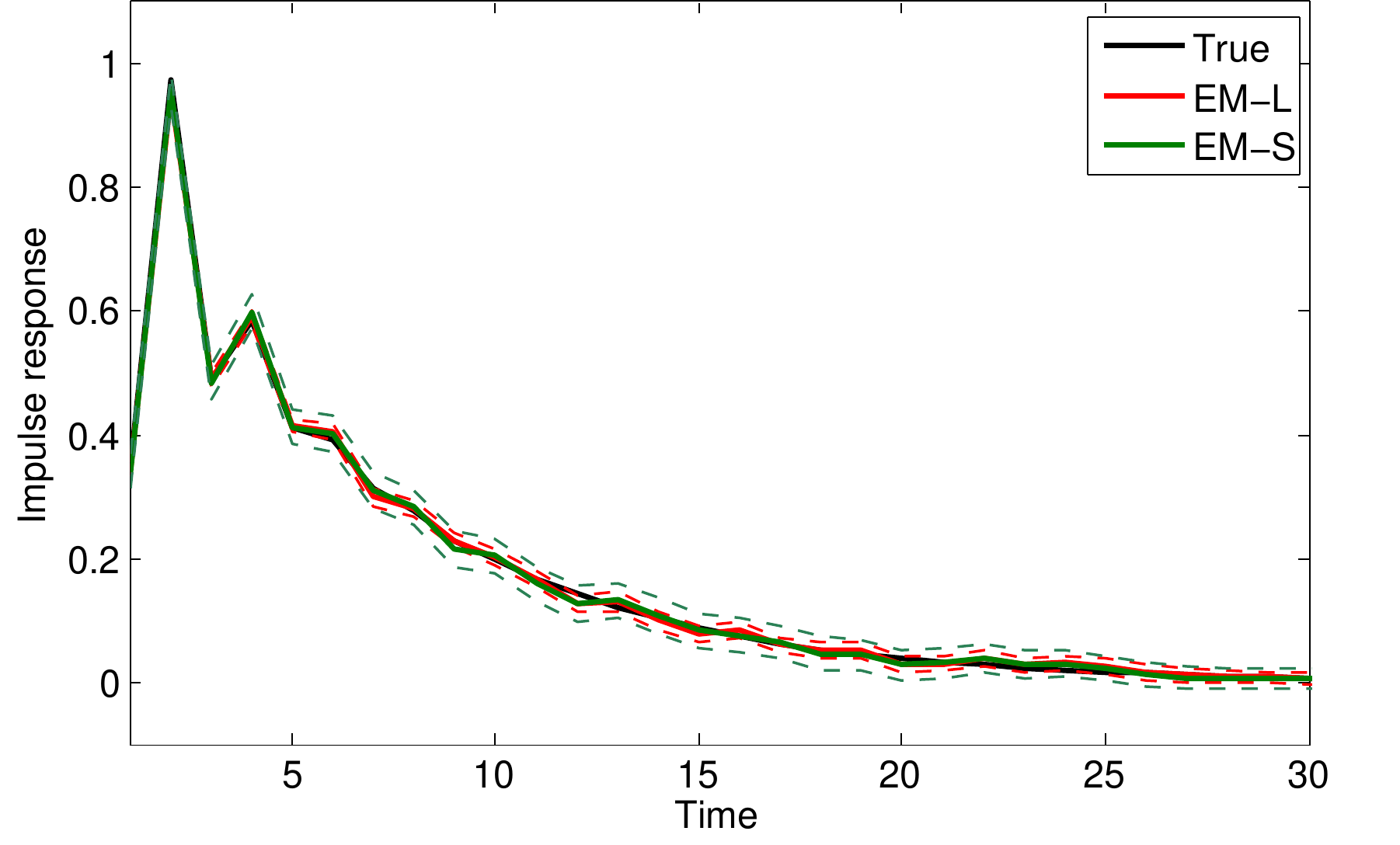}} \\
   \caption{{\bf{Robust solution of the introductory example (Section \ref{sec:numerical5})}}: Impulse response estimates obtained with the proposed method (the dashed lines represent the $99\%$ credibility bounds).} \label{fig:example_proposed}
    %%COMMENT: If possible, use a different line type for each situation (e.g. dashed, dash-dot, solid), in addition to different colors (in case it is printed in black/white).
    %%Also, you may want to make the lines thicker, but this is cosmetic. :)
\end{center}
\end{figure}

We apply the proposed algorithm also to solve the motivating example shown in Section \ref{sec:example}.
Results are visible in Figure \ref{fig:example_proposed}: in comparison with the estimates reported in Figure \ref{fig:example}, the quality of the reconstruction is clearly improved. This can be appreciated also
by inspecting Table 3, % \ref{tab:fits}
which reports the fitting scores of the proposed methods, also comparing them with the non-robust identification method SS-ML applied also in absence of outliers. %to the both the cases where either there are outliers or not.
%Table 3 %\ref{tab:outliers}
%shows the values of each of $\tau_t$ corresponding to those time instants where the outliers are added (i.e., $t = 24,\,39,\,46,\,82,\,94$), comparing them to the median value of all the estimated $\tau_t$. It is clear that the proposed method is able to identify the abnormal measurements. Note that, as $\nu$ increases, the median value of the $\tau_t$ tends to increase as well, while the variance of the outliers tend to shrink.
\begin{table}[h!] \label{tab:fits}
\begin{center}
\begin{tabular}{c|c}
 Method & Fit \% \\
\hline
 EM-S             	& 97.15 \\
 SS-ML (no outliers) &  97.05 \\
 EM-L                 &  95.81 \\
 SS-ML (with outliers) & 70.45 \\

\end{tabular}
\caption{{\bf{Robust solution of the introductory example (Section \ref{sec:numerical5})}}:
Fitting scores of several estimators ranked w.r.t. their performance.}
\end{center}
\end{table}
%% Giulio, it might be best to rank the methods according to score, from best to worst, here.
%
%\begin{table*} \label{tab:outliers}
%\begin{center}
%\begin{tabular}{|c|c|c|c|c|c|c|}
%\hline Method & Median & $t=24$ & $t=39$ & $t=46$ & $t=82$ & $t=94$ \\
%\hline
%\hline EM-L & 8.4 $\times 10^{-4}$ &  1.8771 & 2.4853 & 6.5494 & 6.4728 & 3.1748 \\
%\hline EM-S-2 & 1.93 $\times 10^{-4}$ &  2.7637 & 1.4461 & 16.1567 & 3.9345 & 15.0338 \\
%\hline EM-S-2.25 & 0.0649           &  2.3916 & 1.3245 & 14.9122 & 3.5927 & 13.9444 \\
%\hline EM-S-2.5 & 0.1211          &  2.2679 & 1.3341 & 14.2223& 3.4646 & 13.4026 \\
%\hline EM-S-2.75 & 0.1715          &  2.1711 & 1.3371 & 13.6148& 3.3517 & 12.8957 \\
%\hline EM-S-3 & 0.2177          &  2.0894 & 1.3379 & 13.0696& 3.2520 & 12.4235 \\
%\hline EM-S-5 & 0.4801         &  1.7077 & 1.3285 & 10.0031& 2.6959 & 9.6513 \\
%\hline
%\end{tabular}
%\caption{{\bf{Robust solution of the introductory example (subsection \ref{sec:numerical5})}}:
%capability of detecting outliers variances in the introductory example. The table reports the median values of
%all the estimates of the variances $\tau_t$ and the single estimates of $\tau_t$ associated to the $t$ where outliers
%are present.}
%\end{center}
%\end{table*}

\subsection{Experiment on a real data set} \label{sec:real_experiment}
\begin{figure}[!ht]
\begin{center}
    {\includegraphics[width=8.5cm]{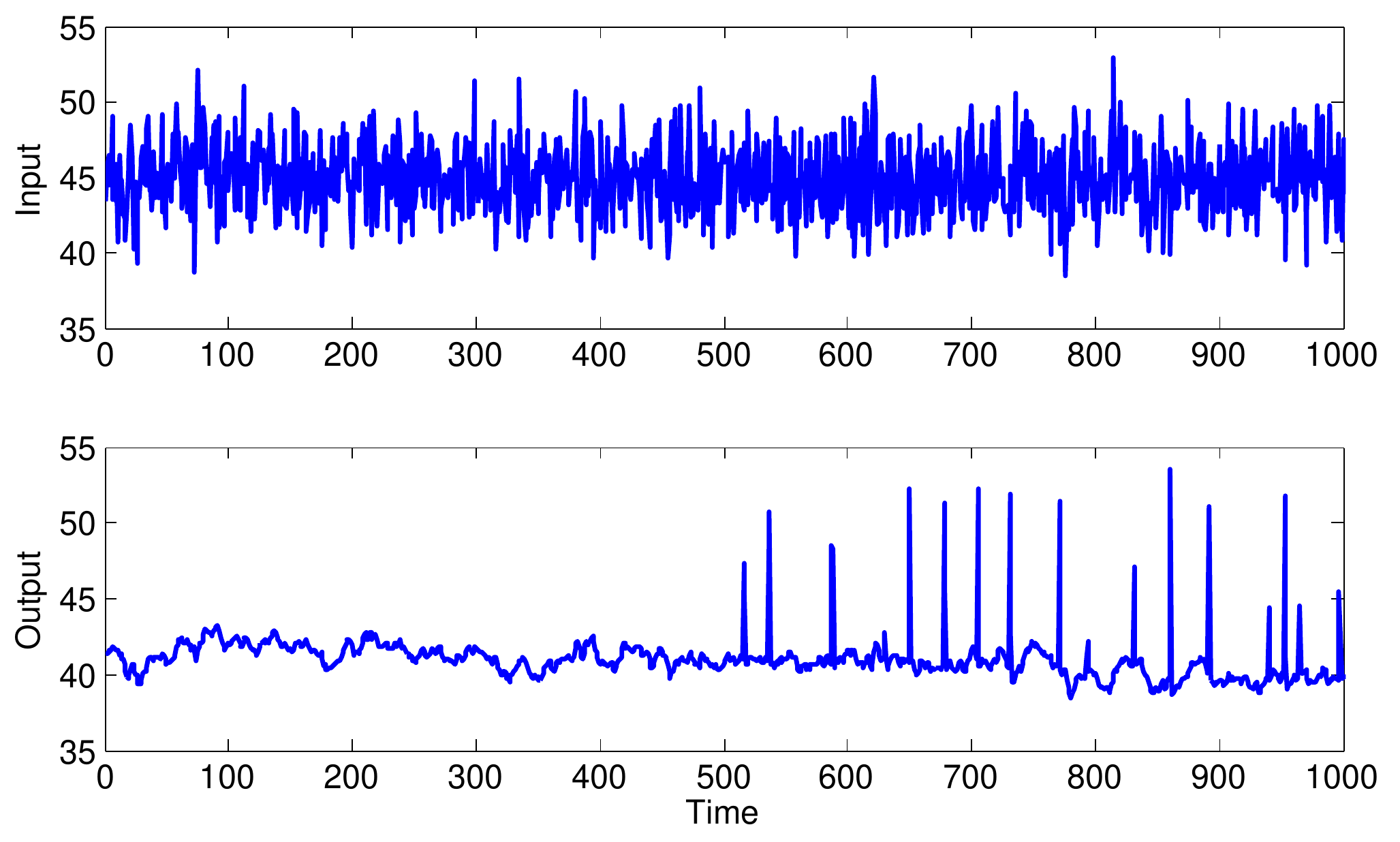}} \\
   \caption{{\bf Experimental data (Section \ref{sec:real_experiment}):} Input and output signals collected during the experiment on the water tank system.} \label{fig:io_data_water}
    %%COMMENT: If possible, use a different line type for each situation (e.g. dashed, dash-dot, solid), in addition to different colors (in case it is printed in black/white).
    %%Also, you may want to make the lines thicker, but this is cosmetic. :)
\end{center}
\end{figure}

We test the proposed method on a real data set. The data are collected from a water tank system. A tank is fed with water by an electric pump. The water is drawn from a lower basin and then flows back to it through a hole in the bottom of the tank. The input of the system is the applied voltage to the
pump and the output is the water level in the tank, measured by a pressure sensor placed at the bottom of the tank (see also \cite{hagg2013robust} for details). Both the input and the output are rescaled in the range $[0,\,100]$, where 100 is the maximum allowed for both the voltage and water level. The input signal consists of samples of the type $u_t = 45 + \sqrt{5}w_t$, with $w_t$ being Gaussian random samples of unit variance. Each input sample is held for 1 second, and one input/output pair is collected each second, for a total of 1000 pairs. The obtained trajectories are depicted in Figure \ref{fig:io_data_water}. As can be seen, the second part of the output signal is corrupted by outliers caused by perturbations of the pressure in the tank (air is occasionally blown in the tank during the experiment, altering the  pressure perceived by the sensor).
Using the methods EM-L, EM-S and SS-ML, we estimate models of the system using the second part of the data set, i.e. $\{y_t,\, u_t\}_{t=501}^{1000}$. The performance of each method is evaluated by computing the fit of the predicted output on the first part of the data set, i.e.
\begin{equation}
FIT_y = 1-\frac{\|y_{\mathrm{test}} - \hat{y}_{\mathrm{test}}\|_2}{\|y_{\mathrm{test}} - \bar y_{\mathrm{test}}\|_2} \,,
\end{equation}
where $y_{\mathrm{test}}$ is the vector containing the output of the test set,  $\bar y_{\mathrm{test}}$ its mean, and
$\hat{y}_{\mathrm{test}} = \hat{U}_{\mathrm{test}} \hat g$ the value predicted by the methods. To get a reference for comparison, we also compute the fit obtained by the method SS-ML when using the test set as training data. Table 4 reports the obtained fits, showing a clear advantage in using the robust methods; in particular, the method EM-S return a fit close to the fit obtained using the test set for identifying the model with the method SS-ML.
\begin{table}[h!] \label{tab:comparing_real_experiment}
\begin{center}
\begin{tabular}{c|c}
 Method & Fit \% on test set  \\
\hline
 SS-ML (estimated using test set) & 70.06 \\
 EM-S & 67.40  \\
 EM-L & 51.81 \\
 SS-ML & 41.49 \\
\end{tabular}
\caption{{\bf{Experiment on a real data set (Section \ref{sec:real_experiment})}}: Fitting scores on the test set.}
\end{center}
\end{table}

\section{Conclusions}
We have proposed a novel regularized identification scheme robust against outliers. The recently proposed nonparametric kernel-based methods
\cite{SS2010}, \cite{ChenOL12} constitute our starting point. These methods use particular kernel matrices, e.g. derived by stable spline kernels,
and Gaussian noise assumptions. This can be a limitation if outliers corrupt the output measurements.
In this paper, we instead model the noise using heavy-tailed pdf's, in particular Laplacian and Student's t.
Exploiting their representation as scale mixture of Gaussians,
we have shown that joint estimation of
noise and kernel hyperparameters can be performed efficiently.
In particular, our robust kernel based method for linear system identification
relies on EM iterations which are essentially all available in closed form.
Numerical results and a real experiment show the effectiveness of the proposed method in contrasting outliers.
%
%The advantage of this scheme compared to the MCMC one proposed before is two-fold:
%\begin{enumerate}
%\item This scheme estimates also an optimal value of $\beta$;
%\item The identification algorithm is much faster.
%\end{enumerate}
%
%An interesting aspect is to model noise using the Student's-t distribution instead of the Laplacian. The scale mixture of Gaussians becomes different, since the $\tau_t$'s become Gamma distributed; this translates into a different optimal value of the $\tau_t$'s in each EM iteration. It is the interesting to assess which model for the noise gives better performances.

\appendix
\section{Appendix}
\subsection{Preliminaries}
First, we show how to compute the E-step of the EM scheme introduced in Section \ref{sec:EM_scheme}, i.e., which form $Q(\theta,\,\hat \theta^{(k)})$ assumes in our problem. The computation of the M-step corresponds to the proof of Theorem \ref{th:min_tau} and is given in the next section.
%the matrices determining mean and variance of $p(g|y,\,\hat \theta^{(k)})$ (see ), computed using the estimate .

\begin{lem} \label{th:Q_theta}
Let $\hat C^{(k)}$ and $\hat P^{(k)}$ be defined as in \eqref{eq:CandP}, computed using the vector $\hat \theta^{(k)}$. Then,
\begin{align} \label{eq:Q_final1}
Q(\theta,\,\hat \theta^{(k)}) &= -\frac{1}{2} \Big(   y^T M(\theta,\,\hat \theta^{(k)}) y +  \log\det \lambda K_\beta  \nonumber \\
 & + \Tr \left\{ \left( U^T \Sigma_v^{-1} U + (\lambda K_\beta)^{-1}\right)\hat P^{(k)} \right\}  \nonumber \\
&  + \sum_{i=1}^N \log \tau_t -2\sum_{i=1}^N  L(\tau_t) \Big) \,,
\end{align}
where
\begin{align} \label{eq:M_theta}
M(\theta,\,\hat \theta^{(k)}) & := \Sigma_v^{-1}(I -2U \hat C^{(k)}) \nonumber \\
    & 	\quad + \hat C^{(k)T} \left( U^T\Sigma_v^{-1}U + (\lambda K_\beta)^{-1}\right)\hat C^{(k)} \,.
\end{align}
\end{lem}

\textbf{Proof: } First, note that $p(y,g|\theta) = p(y|g,\,\theta) p(g|\theta)$,  where
\begin{equation}
p(y|g,\,\theta) \sim \mathcal N \left(Ug,\,\Sigma_v\right)
\end{equation}
and $p(g|\theta)$ is given by \eqref{eq:model_g}.  Hence
\begin{align} \label{eq:log_L}
L(y,g|\theta) + \sum_{t=1}^N  L(\tau_t) & = \nonumber \\
& \!\!\!\!\!\!\!\!\!\!\!\!\!\!\!\!\!\!\!\!\!\!\!\!\!\!\!\!\!\!\!\!\!\!\!\! = -\frac{1}{2} \sum_{t=1}^N \log \tau_t  -\frac{1}{2} (y-Ug)^T \Sigma_v^{-1}(y-Ug)  \nonumber \allowdisplaybreaks[1] \\
& \!\!\!\!\!\!\!\!\!\!\!\!\!\!\!\!\!\!\!\!\!\!\!\!\!\!\!\!\!\! -\frac{1}{2} \log\det \lambda K_\beta -\frac{1}{2} g^T (\lambda K_\beta)^{-1} \! g + \! \sum_{t=1}^N \! L(\tau_t) \nonumber \allowdisplaybreaks[1] \\
 & \!\!\!\!\!\!\!\!\!\!\!\!\!\!\!\!\!\!\!\!\!\!\!\!\!\!\!\!\!\!\!\!\!\!\!\!= -\frac{1}{2} \sum_{t=1}^N \log \tau_t -\frac{1}{2}  y^T \Sigma_v^{-1} y +  y^T\Sigma_v^{-1}Ug \nonumber \allowdisplaybreaks[1]\\
& \!\!\!\!\!\!\!\!\!\!\!\!\!\!\!\!\!\!\!\!\!\!\!\!\!\!\!\! -\frac{1}{2} g^T \left(U^T\Sigma_v^{-1}U + (\lambda K_\beta)^{-1}\right)g  \nonumber\allowdisplaybreaks[1] \\
& \!\!\!\!\!\!\!\!\!\!\!\!\!\!\!\!\!\!\!\!\!\!\!\!\!\!\!\! -\frac{1}{2} \log\det \lambda K_\beta + \sum_{i=1}^N  L(\tau_t) \nonumber \allowdisplaybreaks[1]\\
& \!\!\!\!\!\!\!\!\!\!\!\!\!\!\!\!\!\!\!\!\!\!\!\!\!\!\!\!\!\!\!\!\!\!\!\!:= A + B + C + D + E + F\,.
\end{align}
Now, we have to take the expectation of the above terms with respect to $p(g|y,\,\hat \theta^{(k)})$, which is given by \eqref{eq:pg}. We have the following results  (we make use of the symbol $\mathbb{E}[\cdot]$ to denote such an expectation).
\begin{align} \label{eq:expecatations}
\mathbb{E}[A] &= -\frac{1}{2} \sum_{t=1}^N \log \tau_t \nonumber \\
\mathbb{E}[B] &= -\frac{1}{2}  y^T \Sigma_v^{-1} y \nonumber \\
\mathbb{E}[C] &= y^T\Sigma_v^{-1}U \mathbb{E}[g] =  y^T\Sigma_v^{-1}U \hat C^{(k)} y \nonumber \allowdisplaybreaks[1] \\
\mathbb{E}[D] &=  -\frac{1}{2} \mathbb{E} \left[ \Tr \left\{\left(U^T\Sigma_v^{-1}U + (\lambda K_\beta)^{-1}\right)g g^T \right\} \right] \nonumber \\
              &= -\frac{1}{2}  \Tr \left\{ \left( U^T \Sigma_v^{-1} U + (\lambda K_\beta)^{-1}\right)\hat P^{(k)} \right\}  \nonumber \\
              &\quad - \frac{1}{2} y^T  \hat C^{(k)T} \left( U^T\Sigma_v^{-1}U + (\lambda K_\beta)^{-1}\right)\hat C^{(k)}y \nonumber \\
\mathbb{E}[E] &= -\frac{1}{2} \log\det \lambda K_\beta \nonumber\allowdisplaybreaks[1] \\
\mathbb{E}[F] &= \sum_{t=1}^N  L(\tau_t) \,.
\end{align}
Summing up all these elements we obtain \eqref{eq:Q_final1}.
\hfill $\Box$

%Now, to obtain $\hat \theta^{(k+1)}$ we have to compute the M-step, which consists in maximizing $Q(\theta,\,\hat \theta^{(k)})$. The following result shows that the M-step boils down to a very simple and computationally fast operation.
\subsection{Proof of Theorem \ref{th:min_tau}}
The proof is composed of two parts: (1) We rewrite $Q(\theta,\,\hat \theta^{(k)})$ in a convenient form; (2) we compute the vector $\theta$ maximizing $Q(\theta,\,\hat \theta^{(k)})$.

\emph{Part $(1)$:$\qquad$}We show that the function $Q(\theta,\,\hat \theta^{(k)})$ can be rewritten
\begin{equation} \label{eq:Q_rew_1}
Q(\theta,\,\hat \theta^{(k)})\! = \!-\frac{1}{2} \! \left( \! Q_0(\lambda,\,\beta,\,\hat \theta^{(k)}) + \sum_{t=1}^N Q_t(\tau_t,\,\hat \theta^{(k)}) \! \right) + c,
\end{equation}
where $c$ is a constant,
\begin{align} \label{eq:Q_0}
Q_0(\lambda,\,\beta,\,\hat \theta^{(k)}) & := y^T \hat C^{(k)T} (\lambda K_\beta)^{-1}\hat C^{(k)} y + \log\det \lambda K_\beta \nonumber \\
    &   + \Tr \left\{ (\lambda K_\beta)^{-1} \hat P^{(k)} \right\}
\end{align}
is a function of $\lambda$ and $\beta$ only and, for $t=1,\,\ldots,\,N$,
\begin{equation} \label{eq:Q_tau2}
Q_t(\tau_t,\,\hat \theta^{(k)}) := \hat \varepsilon_t^{(k)} \tau_t^{-1} + \log \tau_t -2 L(\tau_t) \,,
\end{equation}
are $t$ distinct functions, each depending  only on $\tau_t$,  $t=1,\,\ldots,\,N$.

For convenience, we rewrite the matrix in \eqref{eq:M_theta} as $M(\theta,\,\hat \theta^{(k)}) = M_H(\hat \theta^{(k)}) + M_\tau(\hat \theta^{(k)})$, where
\begin{equation}
M_H(\hat \theta^{(k)}) := \hat C^{(k)T} (\lambda K_\beta)^{-1}\hat C^{(k)}
\end{equation}
does not depend on the $\tau_t$ and, conversely,
\begin{align}
M_\tau(\hat \theta^{(k)}) & := \Sigma_v^{-1}(I -2U \hat C^{(k)}) + \hat C^{(k)T} U^T\Sigma_v^{-1}U\hat C^{(k)}
\end{align}
depends only on the $\tau_t$.  Rearranging \eqref{eq:Q_final1}, it follows that
\begin{align} \label{eq:Q_rewritten}
Q(\theta,\,\hat \theta^{(k)}) & = -\frac{1}{2} \Big(   y^T M_H(\hat \theta^{(k)}) y +  \log\det \lambda K_\beta  \nonumber \\
    & \!\!\!\!\!\!\!\!\! + \Tr \left\{ (\lambda K_\beta)^{-1} \hat P^{(k)}\right\} + \Tr \left\{U^T \Sigma_v^{-1} U \hat P^{(k)} \right\}  \nonumber \\
    & \!\!\!\!\!\!\!\!\!  + y^T M_\tau(\hat \theta^{(k)}) y  + \sum_{t=1}^N \log \tau_t -2\sum_{t=1}^N  L(\tau_t) \Big) \,.
\end{align}
where the first part corresponds to $Q_0(\lambda,\,\beta,\,\hat \theta^{(k)})$, defined in \eqref{eq:Q_0}.
Now, let $\hat S^{(k)}$ be as in \eqref{eq:pred_variance}; then
\begin{equation} \label{eq:tau_min1}
\Tr \{U^T \Sigma_v^{-1} U \hat P^{(k)}\} = \Tr \{\hat S^{(k)}  \Sigma_v^{-1}\} = \sum_{t=1}^N \hat s^{(k)}_{tt}\tau_t^{-1} \,,
\end{equation}
where $\hat s^{(k)}_{tt}$ denotes the element of $\hat S^{(k)} $ in position $(t,t)$. Recall also that $\hat y^{(k)} = U \hat C^{(k)} y$; thus
\begin{equation} \label{eq:tau_min2}
y^T  \hat C^{(k)T} U^T\Sigma_v^{-1}U \hat C^{(k)}y = \sum_{t=1}^N \hat y_t^{(k)2} \tau_t^{-1} \,;
\end{equation}
furthermore, since
$
y^T \Sigma_v^{-1} y = \sum_{t=1}^N y_{t}^2\tau_t^{-1}
$
and
\begin{equation} \label{eq:tau_min4}
- 2y^T\Sigma_v^{-1}U \hat C^{(k)} y = -2\sum_{t=1}^N \hat y^{(k)}_{t}y_{t}\tau_t^{-1} \,,
\end{equation}
it follows that
\begin{equation}
y^T M_\tau(\hat \theta^{(k)}) y = \sum_{t=1}^N \left(y_t-\hat y^{(k)}_{t}\right)^2\tau_t^{-1} \,.
\end{equation}
Recalling that $\hat \varepsilon_t^{(k)} = (y_t - \hat y^{(k)}_t)^2 + \hat s^{(k)}_{tt}$, \eqref{eq:Q_tau2} and \eqref{eq:Q_rew_1} follow.

\emph{Part $(2)$:$\qquad$} First, we proceed with the minimization of each $Q_t(\tau_t,\,\hat \theta^{(k)})$, $t = 1,\,\ldots,\,N$; then, we show how to deal with \eqref{eq:Q_0}.

Depending on the noise model adopted, $Q_t(\tau_t,\,\hat \theta^{(k)})$, $t = 1,\,\ldots,\,N$, has two different forms.
\begin{enumerate}
\item If the noise is modeled with the Laplacian density, from \eqref{eq:p_tau_i_Laplacian}  we have $L(\tau_t) = - \frac{1}{\sigma^2}\tau_t + c$, where $c$ is constant so that, for every $t =1,\,\ldots,\,N$
\begin{equation}
Q_t(\tau_t,\,\hat \theta^{(k)}) := \hat  \varepsilon_t^{(k)} \tau_t^{-1} + \log \tau_t +\frac{2}{\sigma^2}\tau_t \,,
\end{equation}
which is minimized by \eqref{eq:min_Laplacian}.
\item If noise is modeled with the Student's t-distribution, from \eqref{eq:p_tau_t_Student}  we have $L(\tau_t) = -\left(\frac{\nu}{2} + 1\right) \log \tau_t - \frac{(\nu-2)\sigma^2}{2\tau_t} + c$, where $c$ is constant so that, for every $t =1,\,\ldots,\,N$
\begin{equation}
\!\!\!\!Q_t(\tau_t,\,\hat \theta^{(k)}) := \left(\hat  \varepsilon_t^{(k)} +  {(\nu-2)\sigma^2} \right) \tau_t^{-1} + (\nu + 3) \log \tau_t  \,.
\end{equation}
The minimizer \eqref{eq:min_Student} follows from solving the above equation.
\end{enumerate}

We now deal with  $Q_0(\lambda,\,\beta,\,\hat \theta^{(k)})$. Its derivative with respect to $\lambda$ is
\begin{equation}
\frac{\partial Q_0}{\partial \lambda} = -\frac{1}{\lambda^2} \left( y^T \hat C^{(k)T} K_\beta^{-1}\hat C^{(k)} y  + \Tr \left\{ K_\beta^{-1} \hat P^{(k)} \right\}\right) + \frac{n}{\lambda} \,,
\end{equation}
which is equal to zero for
\begin{equation} \label{eq:lambda_opt}
\lambda^* = \frac{1}{n} \left(  y^T \hat C^{(k)T} K_\beta^{-1}\hat C^{(k)} y + \Tr \left\{ K_\beta^{-1} \hat P^{(k)} \right\}\right) \,.
\end{equation}
Plugging back such value into $Q_0(\lambda,\,\beta,\,\hat \theta^{(k)})$, one obtains
\begin{align} \label{eq:min_beta_slow}
Q_0(\lambda^*,\,\beta,\,\hat \theta^{(k)}) & =  n\log \left(  y^T \hat C^{(k)T} K_\beta^{-1}\hat C^{(k)} y \right. \nonumber \\
    & \left. + \Tr \left\{ K_\beta^{-1} \hat P^{(k)} \right\}\right) + \log \det K_\beta + c\,,
\end{align}
where $c$ is a constant. Now consider the following factorization of the first order stable spline kernel \cite{carli2014maxentr}:
\begin{equation} \label{eq:factorization_peopeo}
K_\beta = \Delta^{-1} W_\beta \Delta^{-T} \,,
\end{equation}
where $\Delta$ is defined in \eqref{eq:matrix_T} and
\begin{equation}
W_\beta  := (1 - \beta) \diag\left\{\beta,\,\ldots,\,\beta^{n-1},\,\frac{\beta^{n}}{1-\beta}\right\} \,.
\end{equation}
Note that the nonzero elements of $W_\beta$ correspond to the inverse of those of $\eqref{eq:vector_w}$. According to \eqref{eq:transformation1} and \eqref{eq:transformation2}, let $\widehat {\delta g}^{(k)}:= \Delta \hat C^{(k)} y$, $\hat H^{(k)} := \Delta \hat P^{(k)} \Delta^{T}$. Then it can be seen that
\begin{align} \label{eq:Q_after_rewrite}
Q_0(\lambda^*,\,\beta,\,\hat \theta^{(k)}) & = n \log \left( \sum_{i=1}^{n} (\hat h^{(k)}_{ii} + \widehat {\delta g}_i^{(k)2}) w_{\beta,ii}^{-1}\right) \nonumber \\
                                           & +  \sum_{i=1}^{n} \log  w_{\beta,ii} + c\,,
\end{align}
where $\hat h_{ii}^{(k)}$ and $w_{\beta,ii}$ are the $i$-th diagonal elements of $\hat H^{(k)}$ and $W_{\beta}$, respectively, while $ \widehat {\delta g}_i^{(k)}$ is the $i$-th entry of $\widehat {\delta g}^{(k)}$. A further rewriting of \eqref{eq:Q_after_rewrite} yields
\begin{align} \label{eq:Q_after_rewrite_2}
Q(\beta) := Q_0(\lambda^*,\,\beta,\,\hat \theta^{(k)}) & = n \log f(\beta) +  \frac{n(n+1)}{2} \log \beta  \nonumber \\
& \quad + (n-1)\log (1-\beta)+c
\end{align}
where
\begin{equation}
\!f(\beta)\! =\!\! \sum_{i=1}^{n-1} (\hat h^{(k)}_{ii} + \widehat {\delta g}_i^{(k)2}\!) \beta^{1-i} + (\hat h^{(k)}_{nn} + \widehat {\delta g}_n^{(k)2})(1-\beta)\beta^{1-n} ,
\end{equation}
so that \eqref{eq:Q_beta} and \eqref{eq:function_f} are obtained. Using similar arguments, one can see that \eqref{eq:lambda_opt} can be rewritten as \eqref{eq:min_lambda}.

\section*{Acknowledgements}
We thank Niklas Everitt for helping with the experiment on the water tank system.

\bibliographystyle{plain}
\bibliography{biblio}

\end{document}